\documentclass[preprint,amssymb,nofootinbib,amsmath,11pt]{JHEP3}

\JHEPspecialurl{http://jhep.sissa.it/JOURNAL/JHEP3.tar.gz}

\usepackage{epsfig,multicol,amsmath}

\JHEPspecialurl{http://jhep.sissa.it/JOURNAL/JHEP3.tar.gz}
\usepackage{bm}  
\usepackage{amsmath}     
\usepackage{epsfig,epsf}
\usepackage{graphicx}
\usepackage{epstopdf}

\def\beq{\begin{equation}}
\def\eeq{\end{equation}}
\def\bea{\begin{eqnarray}}
\def\eea{\end{eqnarray}}

\def\eq#1{{Eq.~(\ref{#1})}}
\def\fig#1{{Fig.~\ref{#1}}}

\newcommand{\Lb}{\left(}
\newcommand{\Rb}{\right)}

\parskip=\itemsep               

\newcommand{\g}{{\rm g}}

\newcommand{\A}{{\cal A}}

\title{\LARGE \bf D-instantons and multiparticle production in N=4 SYM}
\author{\large  D.E.~Kharzeev$^{\,a}$\,\, and \,\,E.M.~Levin$^{\,a,b}$  \\
a)\,\, Department of Physics, Brookhaven National Laboratory,\\
Upton, New York 11973-5000, USA\\
b)\,  \,Department of Particle Physics, School of Physics and Astronomy\\
Raymond and Beverly Sackler
 Faculty
of Exact Science\\  Tel Aviv University, Tel Aviv, 69978, Israel\\
}

\preprint{BNL-90536-2009-JA\\
TAUP-2904-09\\
{\tt 0910.3355[hep-ph]}\\
\today }

\abstract{
N=4 Super-symmetric Yang-Mills theory (N=4 SYM) in the strong coupling regime has been successfully applied (through the AdS/CFT correspondence)  to the description of strongly coupled plasma which is a multiparticle state.
Yet, the high-energy scattering in the strong coupling limit of N=4 SYM is purely elastic,
so this multiparticle final state can never be produced:
this is because in this limit the theory is dual to
weak supergravity, and the dominant interaction is the elastic graviton exchange.
Here we propose a resolution of this dilemma by considering the contribution of
D-instantons in $AdS_5$ bulk space to the scattering amplitude. We argue that D-instantons coupled to dilatons and axions are responsible for multiparticle production in strongly coupled N=4 SYM, and the corresponding cross section increases with energy. We evaluate the intercept and the slope of the corresponding Pomeron trajectory in terms of the typical size of
the D-instanton, and argue that the resulting physical picture may resemble the real world.
 }


\begin{document}

\section{Introduction}
N=4 Super-symmetric Yang-Mills (SYM) theory does not possess the properties of confinement and asymptotic freedom. Therefore, it cannot be considered a theory adequately describing the strong interactions in the real world. However, this theory 
has two unique attractive features. First,
 N=4 SYM at weak coupling has some properties similar to perturbative QCD (see Refs.
\cite{POST,BFKL4}):  the Operator Product Expansion (OPE)  and the linear evolution equations for Deep-Inelastic Scattering (DIS) are valid, as well as the BFKL equation for the high energy scattering amplitude.
The high energy amplitude reaches the unitarity limit, or the "black disc" regime in which half of the cross section stems from the elastic scattering and half from the inelastic processes of multiparticle production.
Second, in the strong coupling regime N=4 SYM can be reduced to weak classical gravity in the $AdS_5$ bulk space through the AdS/CFT correspondence \cite{AdS-CFT}, and can be solved analytically. Therefore, for physical systems for which  neither confinement nor running  coupling are essential, the N=4 SYM theory can provide guidance in the strong coupling regime. 
In particular, it may lead to a better understanding of phenomena in strongly coupled QCD matter.  Indeed,  some of the main successes of N=4 SYM  to date have been achieved in the description of     the quark-qluon plasma at strong coupling  \cite{KSS,HKKKY,MUN4,PLASMAN4}.

\FIGURE[h]{
\begin{minipage}{80mm}
\centerline{\epsfig{file=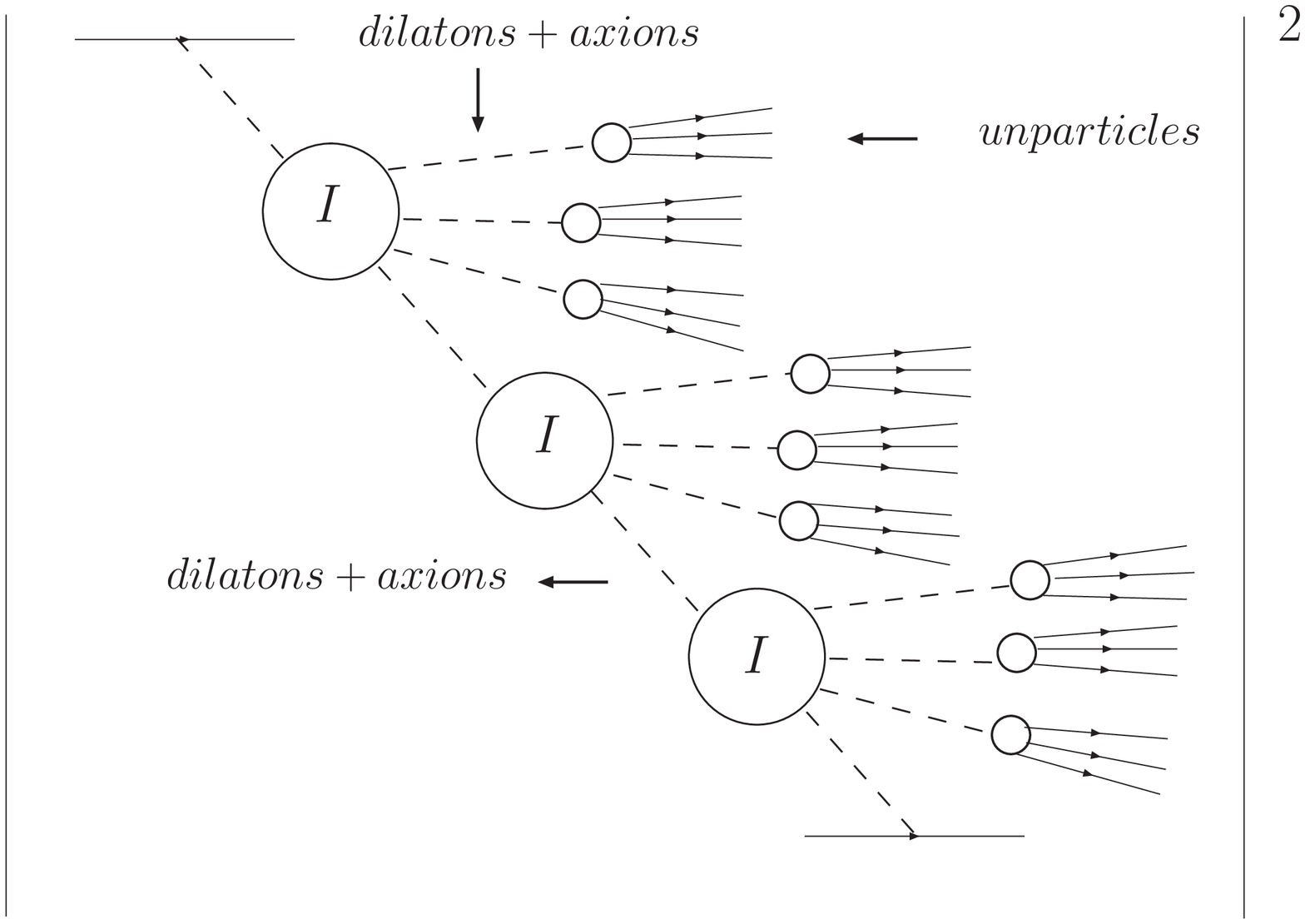,width=70mm}}
\caption{ The process of multiparticle production in N=4 SYM}
\end{minipage}
\label{instgen}}

However, the physical picture of the high-energy collision in the strong coupling regime of N=4 SYM turns out
to be completely different\cite{BST1,BST2,MHI,BST3,COCO,BEPI,LMKS} from the one at weak coupling:  the 
multiparticle production in this case is apparently unimportant and the main contribution stems from elastic and quasi-elastic (diffractive)  processes when the target (proton or a nucleus) either remains intact, or is slightly excited. This picture contradicts both 
the QCD expectations\cite{GLR,MUQI,MV,BFKL,BK,JIMWLK} and the available experimental data.

Since the creation of quark-gluon plasma necessarily proceeds through the multiparticle production, we face a rather controversial situation. 
Indeed, the multiparticle systems which we hope are adequately described by the strongly coupled N=4 SYM are not  produced in this theory.  The reason for this is simple: the main contribution to the scattering at high energies stems from the exchange of the graviton which has the spin equal to 2 and generates large but {\it real} scattering amplitude. The unitarity constraint for the scattering amplitude has the form
\beq \label{I1}
2\,\mbox{Im} \A\Lb s,b\Rb\,\,\,=\,\,\,| A\Lb s,b\Rb|^2\,\,\,+\,\,{\cal O}\Lb \frac{2}{\sqrt{\lambda}}\Rb
\eeq
where $\lambda = g^2_{YM}\,N_c=\,4\pi\,N_c\, g_s$. We consider the regime where the number of colors is large
$N_c \,\gg\,1$,  the SYM coupling is strong $\lambda \gg 1$, but the string coupling constant is small $g_s \ll 1$. This allows for a perturbative dual description in terms of the graviton interactions.
Geometrically, the radius of the $AdS_5$ space $R$  is related to the coupling constant by $R^2 = \sqrt{\lambda}\ \alpha^\prime$ ($2 \pi \alpha^\prime$ is the inverse of the string tension) so that large values of $\lambda \gg 1$ correspond to the $AdS_5$ space of large radius and small curvature for which the perturbative gravitational description is valid.

\FIGURE[h]{
\centerline{\epsfig{file=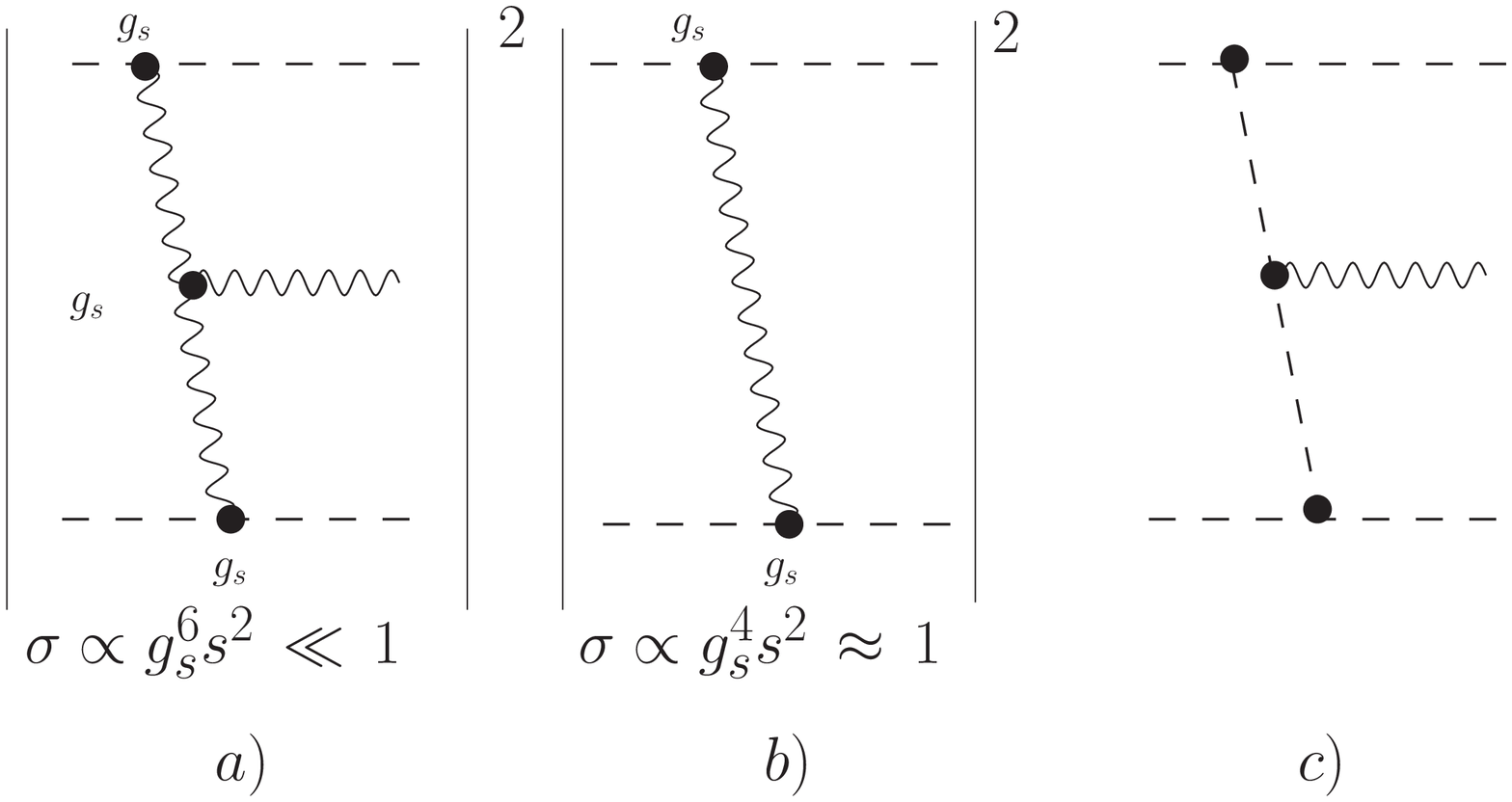,width=100mm,height=40mm}}
\caption{ The graviton production in N=4 SYM. Gravitons are denoted by wavy lines while dilatons by dashed ones.}
\label{grprod}}

The unitarity constraint of \eq{I1} follows from the eikonal formulae (see Refs.\cite{BST2,MHI,COCO,LMKS})
that sum the main contributions of the order of $ \Lb g_s \,\alpha'\,s\Rb^n$ and  describe the scattering amplitude in the kinematic region where $g_s\, \alpha'\,s\,\,\approx \,\,1$. The graviton production  in the central rapidity region  leads to the cross section which is proportional to $g^6 \,\alpha'^2\,s^2 $ (see \fig{grprod}-a ) 
and is much smaller than the eikonal diagram of \fig{grprod}-b describing elastic scattering. The estimates for the smallness of the multiparticle production cross section in \eq{I1} follow from the fact that the reggeized graviton has a trajectory with the intercept $j_0 = 2 -2/\sqrt{\lambda}$ (see Ref. \cite{BST1}).

Therefore at $g_s \ll1$ the emission of gravitons is suppressed and, at first sight, we are doomed to a small multiparticle production cross section that follows from \eq{I1}.  However, this conclusion is premature since the production of a large number of particles ($n \gg 1$) 
may not be adequately described by the lowest order perturbation theory. Instead in this case 
we can rely on the classical  (semi-classical)
approach. Indeed, the phase--number uncertainty principle $\Delta n \,\Delta \phi \geq 1$ implies that the variance of the phase $\phi$ of the field describing the large average number $\bar{n}$ of radiated quanta should be small: $\Delta \phi^2 \,\propto 1/{\bar{n}} \,\ll\,1$, so the field is semi-classical. 

The deep connection between the classical solutions and the multi-particle production has been 
realized long time ago \cite{Dyson:1952tj,LN,Brezin,Parisi}. Let us briefly remind the main ideas underlying this connection. Consider first a perturbative expansion of some amplitude $F$ in powers of the coupling constant $e$: $F(e^2) = a_0 + a_2 e^2 + a_4 e^4 + ...$. In the famous paper \cite{Dyson:1952tj} Dyson has argued that this expansion cannot be convergent: if this series were convergent, $F(e^2)$ would be an analytic function of $e$ at $e=0$, and so $F(-e^2)$ would also have a convergent expansion at sufficiently small $e$. However, the negative coupling constant $e^2 < 0$ corresponds to the world where the like charges attract each other. The vacuum in this world is unstable with respect to the spontaneous creation of very large numbers of particles: the vacuum with electrons in one region of space and positrons in another has a lower energy than an "empty" vacuum.  Dyson argued that the divergence of the power series would manifest itself at high orders $\sim 1/e^2$ and would physically describe the virtual processes of multi-particle production.

Lipatov \cite{LN} has demonstrated explicitly that the degree of divergence of the perturbative expansion in scalar theories with dimensionless negative coupling is determined by spherically-symmetrical
solutions of the corresponding classical equations in Euclidean space. For the same reason, classical solutions also dominate large orders of perturbation series if the expansion is performed around an unstable vacuum \cite{Brezin}.  The action of the classical Euclidean solution determines the probability of tunneling from the unstable to the stable vacuum. These methods have been applied to obtain the asymptotic estimates of high orders of perturbation theory in scalar electrodynamics \cite{Parisi}
and for a Yang-Mills field coupled to a scalar field \cite{LN1}.     

In QCD, we are dealing with the situation in which the perturbative expansion is made around 
an unstable vacuum -- the structure of the true QCD vacuum state is not reproduced by perturbation theory. Therefore, we expect that the classical Euclidean solutions of QCD will contribute to the scattering amplitudes, and will describe the processes of multi-particle production. The classical Euclidean solutions -- instantons \cite{Belavin:1975fg} -- have been found in QCD, as well as in   N=4 SYM (see reviews \cite{INSTSH,INSTD,INSTN4} and references therein). It has been demonstrated that the instanton "ladders" lead to large multiparticle production cross section in QCD \cite{KKL} at high energy 
(for instanton-induced effects in high-energy scattering, see also \cite{Ringwald:1998ek,Balitsky:1992vs,Shuryak:2000df,INSTD,Dorokhov:2004fb,Kochelev:2009rf} and references therein).

All of the instanton-based field-theoretical treatments are restricted to the domain of weak coupling. However it appears that instantons also play an important role in supergravity that is related through the AdS/CFT correspondence to the strong coupling regime of N=4 SYM. 
String theory (that reduces to supergravity at weak string coupling $g_s$) possesses solitonic D-branes that are the classical solutions to the supergravity equations of motion. If the Euclidean world-volume of a brane is wrapped around an appropriate compact manifold, we obtain a so-called D-instanton \cite{Gibbons:1995vg,Green:1997tv} that induces a non-perturbative gravitational interaction involving the axion and dilaton fields (see review \cite{BVN} and the next section).

The main idea of this paper is to evaluate the multiparticle production that originates from the D-instantons in weak gravity.  We show that this production is not only large but in the wide region of large $g_s\,N_c$ 
generates the total cross section that increases as a function of energy.  We argue that the main contribution to these processes stems from so-called multiperipheral diagrams of \fig{instgen} that we will evaluate in N=4\, SYM.

The paper is organized as follows.
 We introduce all notations and give a general discussion of the instanton contributions in weakly coupled supergravity in the next section. In section 3 we calculate the diagrams of \fig{instgen}-type and  discuss the energy dependence of the cross section.
We show that in spite of the fact that the exchange of scalar particles leads to $\sigma \,\propto 1/s^2$ the sum of these diagrams gives $\sigma \,\propto\, s^{\Delta -2}$ with $\Delta > 2$ in the wide range of strong coupling $\lambda$.

\section{N=4\,SYM in the strong coupling regime: general approach}
\subsection{Supergravity (boson sector)}

N=4 SYM at strong coupling is related via the AdS/CFT correspondence \cite{AdS-CFT} to 
the supergravity in ten dimensions with the following action (in Euclidean formulation) \cite{BVN}\footnote{The action of \eq{SG1} is written for ten dimrensional flat space (${\bf {\cal R}}10$). For $AdS_5$ the expression is more complicated in its gravity part but the instanton contribution remains to be the same (see Ref. \cite{BVN}).}:
\beq \label{SG1}
-\,S^{\mbox{boson}}_{\mbox{E}}\,\,\,=\,\,\,\int\,d^{10} x \,\sqrt{g} \,\{{\cal R}\,\,-\,\, (\partial_\mu \phi)(\partial^\mu \phi)
\,\,+\,\,\frac{1}{2}\,e^{2 \phi}\,  (\partial_\mu a)(\partial^\mu a)\}
\eeq
The action of \eq{SG1} describes the interaction of bosonic fields that are given by ten-dimensional metric $g_{\mu \nu}$ (gravitons), the dilaton $\phi$ and the axion $a$ (that are scalar and pseudoscalar, respectively).
In \eq{SG1} ${\cal R}$ is constructed from the Ricci tensor and describes the graviton interactions, while the interaction of the dilatons and axions is shown explicitly.

The first message from \eq{SG1} is that graviton cannot be produced due to interaction of two dilatons (see diagram of \fig{grprod}-c which is equal to zero).
 The second one is that the large multiparticle production can originate  only from the dilaton and axion interactions. As we have discussed in the Introduction, this multiparticle production can be essential only if the classical solution for the dilatons and axions exists.

The equation of motion for the action of \eq{SG1} has the following form:
\beq \label{SG2}
{\cal R}_{\mu \nu}\,=\,\frac{1}{2}\,(\partial_\mu \phi)(\partial_\nu \phi)\,-\,\frac{1}{2}\,e^{2 \phi}\,  (\partial_\mu a)(\partial_\nu a);\,\,\,\,\,\,\,\nabla_\mu\Lb e^{2 \phi}\,\partial^\mu a\Rb\,=\,0;\,\,\,\,\,\,\,\nabla^2 \phi\,+\,e^{2 \phi}\,  (\partial_\mu a)(\partial^\mu a)\,=\,0
\eeq
\begin{boldmath}
\subsection{$AdS_5 \times S^5$ space}
\end{boldmath}
We start with some basic facts about $ AdS_5$ which is defined as the hypersurface embedded in six-dimensional flat space-time by the equation  (with d=4)

\beq \label{SG3}
 - Y^2_{-1} \,+\,Y^2_0 \,+\,\sum^d_{I=1} \,X^2_i= - L^2
\eeq
with curvature $R=- d(d -1) L^2\,=\,-12 L^2$. Introducing new coordinates

\beq \label{SG4}
x_i\,\,=\,\,\frac{L\,X_i}{Y_0 + Y_{-1}}\,;\qquad z\,\,=\,\,\frac{L^2}{Y_0 + Y_{-1}}\,;
\eeq
we reduce the introduced metric to the following form

\beq \label{SG5}
d s^2\,\,=\,\,\frac{L^2}{z^2}\,\Lb \,d z^2\,\,+\,\,\sum^d_{i=1}  d x^2_i \Rb\,=\,\frac{L^2}{z^2}\,\Lb \,d z^2\,+\,d \vec{x}^2 \Rb
\eeq
where $\vec{x}$ is four dimensional vector.  Using \eq{SG5} we can construct the invariant volume element of $AdS_5$, namely
\beq \label{INVVO}
\int\,\,d^4x\,d z\,\sqrt{g}\,\,=\,\,L^5\,\int\,d^4 x \frac{d z}{z^5}
\eeq

The metric in the full ten-dimensional  $AdS_5 \times S^5$ space is
\beq \label{SG6}
d s^2\,\,=\,\,\frac{L^2}{z^2}\,\Lb d x_\mu \,dx_\mu\,\,+\,\,d y_i \,d y_i \Rb\,\,\mbox{where}\,\,\, \mu\,=\,1, \dots , 4\; \,\,\mbox{and}\,\, i\,=\,5,\dots ,10;\,\,\,\,\,z^2\,\,=\,\,y_i y_i ;
\eeq
\subsection{D-instanton}
The classical (instanton) solution can be obtained from \eq{SG2} by setting ${\cal R} = 0$. The field equations then become
\beq \label{SG7}
\,\frac{1}{2}\,(\partial_\mu \phi)(\partial_\nu \phi)\,\,=\,\,\frac{1}{2}\,e^{2 \phi}\,  (\partial_\mu a)(\partial_\nu a);\,\,\,\,\,\,\,\nabla_\mu\Lb e^{2 \phi}\,\partial^\mu a\Rb\,=\,0;\,\,\,\,\,\,\,\nabla^2 \phi\,+\,e^{2 \phi}\,  (\partial_\mu a)(\partial^\mu a)\,=\,0
\eeq
where all indices are contracted with the ten-dimensional metric of \eq{SG6}.

It is easy to see that the solution for the axion field equation is
\beq \label{SG8}
a\,\,-\,\,a_{\infty}\,\,=\,\,\pm \Lb e^{ - \phi} \,-\,e^{- \phi_{\infty}}\Rb
\eeq
where both $a_{\infty}$ and $\phi_{\infty} $ are the integration constant for the fields at infinity.

For the dilaton fields we obtain the equation
\beq \label{SG9}
g^{\mu \nu} \nabla_\mu \partial_\nu e^{ \phi}\,\,=\,\,\frac{1}{\sqrt{g}}\partial_\mu\Lb \sqrt{g}\,g^{\mu \nu}\,\partial_\nu e^{\phi}\Rb\,\,\,=\,\,\,0
\eeq
Comparing with the equation for the propagator of the scalar massless field in  $AdS_5$ \cite{GRF4M} one can see that
\beq \label{SG10}
e^{\phi}\,\,=\,\,e^{\phi_{\infty}} \,\,+\,\,G\Lb\vec{x}_0 , z_0;\vec{x},z\Rb\,\,=\,\,g_s + G\Lb\vec{x}_0 , z_0;\vec{x},z\Rb
\eeq
where $G$ is a propagator from the point of the location of the instanton $(\vec{x}_0,z)$ to the point of observation $(\vec x,z)$. \eq{SG8} gives the instanton solution for the axion field.
The general form of $G\Lb\vec{x}_0 , z_0;\vec{x},z\Rb$ can be found in Ref \cite{GRF4M}. Comparing \eq{SG10} with the instanton field in the weak coupling regime we see that $z_0$ plays a role of the instanton size while $\vec{x}_0$ is the position of the instanton.
The invariant volume element of $AdS_5$ (see \eq{INVVO}) gives us the distribution function $n(z)$ of the instanton with respect to its size\footnote{Starting from this formula and below we assume that $L =1$ to make all expressions simpler. In the final answer we will put back the needed power of $L$.}
, $n(z) \propto 1/z^5$.
The contribution of the D-instanton to the action is equal to
\beq \label{SG11}
S_{\mbox{D-inst}}\,\,=\,\,\frac{2 \pi}{g_s}\,\,=\,\,E_{sph}\,t\,\,\equiv\,\,\kappa
\eeq
and can be obtained by direct integration of the field (see \eq{SG10} and Ref.\cite{INSTD}). The typical time $t$ in \eq{SG11} turns out to be $t = z$ ($z$ is the instanton size). Therefore, the estimate for the typical sphaleron energy for the instanton  is
\beq \label{ESPH}
E_{sph}\,\,=\,\,\frac{2 \pi}{g_s\,z}\,\,=\,\,\frac{\kappa}{z}
\eeq
\section{D-instantons and multiparticle production}
\subsection{The general form of the multiperipheral (ladder) diagram}

~

\FIGURE[h]{\begin{minipage}{95mm}
\centerline{\epsfig{file=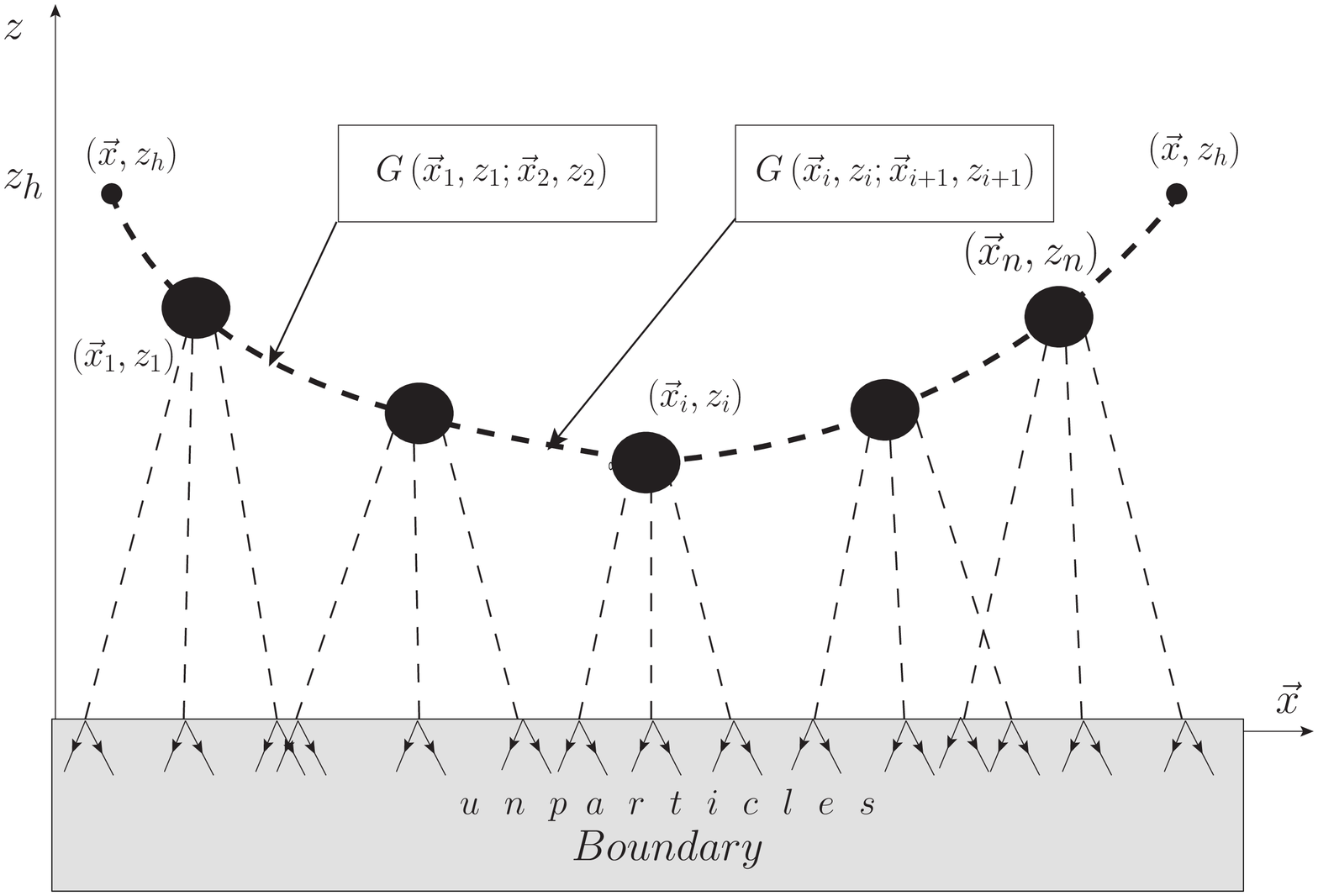,width=90mm}}
\caption{ The amplitude  of multiparticle production in N=4 SYM in coordinate representation. Dashed lines denote the dilatons and axions, the arrows stand for produced unparticles.}
\end{minipage}
\label{ladcr}}

The multiperipheral diagram of \fig{instgen} can be written in more detailed form using \fig{ladcr}. This figure shows the ingredients that we need to calculate the amplitude: the correlation function \\
$\langle \phi_{inst}(\vec{x}_i,z_i)\,
\phi_{inst}(\vec{x}_{i + 1},z_{i + 1})\rangle$ which is equal to the propagator of the scalar particle in our theory $G\Lb
\vec{x}_{i + 1},z_{i + 1};\vec{x}_{i},z_{i }\Rb$ ; the vertex for production of the $n$-particles $\Gamma\Lb 2 \to n\Rb$ from one instanton; and the probability to find one instanton which is equal to
\beq \label{MP1}
\int\,\frac{d z_i}{z_i^5}\,d^4 x_i\,e^{-\frac{2 \pi}{g_s}}\,\,\equiv \int\,\frac{d z_i}{z_i^5}\,d^4 x_i\,e^{- \kappa}
\eeq

The amplitude of the diagram in \fig{ladcr} has the following form in the coordinate representation
\beq \label{MP2}
A\Lb \vec{x},z_h;\vec{x}_{n + 1},z_h\Rb\,\,\,=
\eeq
$$\,\,\,\int\,G\Lb \vec{x},z_h; \vec{x}_1,z_1\Rb \prod^n_{i = 1}\,\frac{d z_i}{z^5_i}\,d^4\,x_i \,e^{ - \frac{2\pi}{g_s}}\,
$$
$$\Gamma( 2 \to\,n_i)\,G\Lb \vec{x}_i,z_i; \vec{x}_{i +1},z_{i + 1}\Rb
$$
where the vertex $\Gamma\Lb 2 \to n_i\Rb$ can be calculated from the Green's function
\bea \label{MP3}
&&G\Lb \vec{x}_k,z_k| k_1,\dots, k_{n_i}\Rb =\\
&&\,\,\langle \phi_{inst}( \vec{x}_k,z_k|\vec{x},z) \, \phi_{inst}( \vec{x}_i,z_i|\vec{x}',z')\,\,\nonumber\\
&&\prod_{i=1}\, \phi_{inst}( \vec{x}_k,z_k|\vec{r}_i,z \to 0)\rangle;\nonumber
\eea
$(\vec{r}_i,z\to 0)$ are the coordinates of the produced particles and $\langle \dots\rangle$ denotes the averaging  over the position of the instanton. The produced particles will be detected at the boundary at large values of $\vec{x}_i$\footnote{We want to recall that $\vec{x}$ denotes the four-vector.}

The general rules for finding the vertex from such kind of the Green's function is the following (see for example Ref. \cite{KKL}): go to momentum representation for produced particle, multiply by inverse propagator $G^{-1}(k_i)$ , in a such amputated Green's function put particles on the mass shell ( say $ k^2_i=0$, as in the case of gluons).
However, we can find the sum over all produced particles (quarks and gluons) in a more elegant way by just using the fact that we need to calculate the cross section which is proportional to $A\Lb \vec{x},z_h;\vec{x}_{n + 1},z_h\Rb\times
A^*\Lb \vec{x},z_h;\vec{x}_{n + 1},z_h\Rb$ and, therefore, what we actually need to know is the Green's function
$\langle 0 | \phi_{inst}( \vec{x}_k,z_k|\vec{r}_i,z \to 0)\,\phi_{inst}( \vec{\bar{x}}_k,\bar{z}_k|\vec{\bar{r}}_i,z \to 0) |0\rangle$ where $\bar{z}_k,\bar{x}_k$ and $\bar{r}$ are coordinates in the complex conjugated amplitude.
The explicit form for $ \phi_{inst}( \vec{x}_k,z_k|\vec{r}_i,z \to 0)$ follows directly from Witten's formula \cite{WI}:
\beq \label{MP4}
\phi_{inst}( \vec{x}_k,z_k|\vec{r}_i,z \to 0)\,\,=\,\,\frac{2 \pi}{g_s}\,\frac{\Gamma(4)}{\pi^2 \Gamma(2)}\frac{z^4_k}{\Lb z^2_k\,\,+\,(\vec{r}_i - \vec{x}_k)^2\Rb^4}\,\,\xrightarrow{r_i \gg x_k}\,\frac{2 \pi}{g_s}\,\,\frac{\Gamma(4)}{\pi^2 \Gamma(2)}\frac{z^4_k}{r^8_i}
\eeq 
We are interested only in the fields at large distances since we consider the scattering amplitude and produced particles can be measured only at large distances.  Thus the Green's function\\
 $\langle 0 | \phi_{inst}( \vec{x}_k,z_k|\vec{r}_i,z \to 0)\,\phi_{inst}( \vec{\bar{x}}_k,\bar{z}_k|\vec{\bar{r}}_i,z  \to 0) |0\rangle$ can be reduced to $\langle 0 | \phi_{inst}(\vec{r}_i)\,\phi_{inst}(\vec{\bar{r}}_i) |0\rangle$ which can be evaluated using the idea of unparticles suggested in Ref. \cite{HG}. Namely,
\beq \label{MP5}
\langle 0 |\, \phi_{inst}(\vec{r}_i)\,\phi_{inst}(\vec{\bar{r}}_i)\, |0\rangle\,\,=\,\,\int\,e^{-i \vec{k}\cdot (\vec{r} - \vec{\bar{r}})}\,\,|\langle 0|\phi |k\rangle|^2\,\rho(k^2)\,\frac{d^4 k}{( 2 \pi)^4}
\eeq
where $|k\rangle$ is a state of the unparticle with the four-momentum $k_\mu$ produced from the vacuum by $\phi_{inst}$.  Due to the scale invariance
\beq \label{MP6}
|\langle 0|\,\phi \,|k\rangle|^2\,\rho(k^2)\,\,=\,\,A^2_{d = 4} \,\Theta(k_0)\,\Theta(k^2)\,(k^2)^{d - 2}
\eeq
\eq{MP6} sums the emission of quarks and gluons in the most economic way demonstrating one of the best features of N=4 SYM approach. Using the dispersion relation one obtains from \eq{MP6}  that the propagator of the unparticle is equal to $D_{unparticle}(k^2)=k^4\ln(- k^2)$.  We can easily find out the coefficient $A_4$ in \eq{MP6}
using the following equation
\beq \label{MP7}
A_4\,\,=\,\,\frac{1}{D_{unparticle}(k^2)}\,\int\,d^4 r\,\, e^{i \vec{k}\cdot \vec{r}}\,\phi_{inst}(z;r)\,\,=\,\,\frac{2 \pi}{g_s}\, \frac{z^4}{4\,4!}
\eeq
where $z$ is the instanton size.

Armed with \eq{MP5} - \eq{MP7} we can explain our pictures in \fig{instgen} and \fig{ladcr}. The dashed line in these figures denote the dilaton(axion) fields induced by the instanton.  Each of such fields creates on the boundary a cluster of produced particles (gluons and quarks); this cluster  can be described as an unparticle. Using \eq{MP5} we can 
calculate the total inelastic cross section by  just summing over the number of  unparticles without deciphering the content of this cluster in terms of quarks and gluons.

\subsection{Summing over produced unparticles}
From \eq{MP3} one can see that we necoordinateed to sum over produced unparticles since the vertex $\Gamma(2  \to n) \propto \phi^n_{inst}$. For the production of unparticles with fixed total mass $q$ we need to find
\beq \label{MP8}
\Sigma_n(q)\,\,\,\equiv\,\,\,n!\,\langle 0 |\, \phi_{inst}(\vec{r}_i)\,\phi_{inst}(\vec{\bar{r}}_i)\, |0\rangle^n\,\Longrightarrow\,\,
(2 \pi)^4 \,\delta^{(4)} \Lb q - \sum^n_{i=1} k_i\Rb\,n!\,\prod^n_{i=1} A^{2 }_4\,\,\Theta(k_{0,i})\,\Theta(k^2_i)\,(k_i^2)^{2}\,\frac{d^4 k_i}{( 2 \pi)^4}
\eeq
The factor $n!$ takes into account the possibility  that two arbitrary $\phi_{inst}$ can enter in  $ \langle 0 |\, \phi_{inst}(\vec{r}_i)\,\phi_{inst}(\vec{\bar{r}}_i)\, |0\rangle$. Replacing
\beq \label{MP91}
(2 \pi)^4 \,\delta^{(4)} \Lb q - \sum^n_{i=1} k_i\Rb\,\,\to\,\,\int\,d^4 x\,\exp\Lb i\, (\vec{q} - \sum \vec{k}_i)\cdot \vec{x}\Rb
\eeq
we reduce \eq{MP8} to
\beq  \label{MP9}
\Sigma_n(q)\,\,\,=\,\,A^{2n}_4\,n!\,\int\,d^4 x\,e^{i \vec{q}\cdot\vec{x}}\,\,\left\{\int\,\frac{d^4 k}{(2 \pi)^4}\,(k^2)^2\,\Theta(k_0)\,\Theta(k^2)\,e^{-i \vec{k} \cdot\,\vec{x}}\right\}^n
\eeq
Using the identity $\Box_x \exp(- i \vec{k}\cdot \vec{x})  = - k^2\,\exp(- i \vec{k}\cdot \vec{x})$ we can rewrite
$\{\dots\}$ in the following way
\beq \label{MP10}
\left\{\dots\right\}\,\,\,\equiv\,\,\Box^2_x\,\,{\cal I}(x)\,\,\,\,\mbox{with}\,\,\,\,{\cal I}(x)\,=\,\int ^{\infty}_0 \frac{d k_0}{2\pi} \,e^{-k_0\,x_0}\,\int^{k_0}_0\,e^{-i\,|k|\,|x|\,\cos\theta}\,d \cos\theta\,\frac{|k|^2 d|k|}{( 2 \pi)^2}
\eeq
where $|k|$ and $|x|$ are the length of the three-vectors $k$ and $x$ repectively.

After integration over $\cos\theta$ and $|k|$ we obtain
\beq \label{MP11}
{\cal I}(x)\,\,=\,\ \int ^{\infty}_0 \frac{d k_0}{2\pi^3} \,e^{-k_0\,x_0}\,\frac{k_0\,|x|\,\cos(k_0\,|x|) \,-\,\sin(k_o|x|)}{|x|^3}
\eeq
Integrating over $k_0$ and calculating $\Box^2_x {\cal I}(x)$ we get
\beq \label{MP12}
\Box^2_x\,\, {\cal I}(x)\,\,=\,\,\frac{24 \cdot 8}{\pi^3}\,\frac{1}{(x_0^2 - |x|^2)^4}
\eeq
In all calculations above we assumed that $q_\mu = (q,0,0,0)$. Using  \eq{MP12} we have for $\Sigma_n(q)$ the following expression
\bea \label{MP13}
\Sigma_n(q)\,\,\,&=&\,\,A^{2n}_4\,n!\,\Lb \frac{8\cdot 24}{\pi^3}\Rb^n\,\int\,d^4 x\,\,e^{i\,q\,x_0}\,\frac{1}{x^{8\,n}}\,\ \nonumber\\
\,\,&=&\,\,- i\,A^{2n}_4\,n!\,\Lb \frac{8\cdot 24}{\pi^3}\Rb^n\,4\,\pi\,\int^{+\infty}_{-\infty}\,\,e^{i\,q\,x_0}\,d x_0\,\frac{1}{x_0^{8 n - 3}}\,\,
\frac{\pi^{3/2}\,\Gamma\Lb 4\,n - 3/2\Rb}{\Gamma\Lb 4\,n\Rb} \nonumber\\
\,\,&=&\,\,A^{2n}_4\,n!\,\Lb \frac{8\cdot  24}{\pi^3}\Rb^n\,\frac{\,\pi^{5/2}\,\Gamma\Lb 4\,n - 3/2\Rb}{\Gamma\Lb 4\,n + 1\Rb\,\,\Gamma\Lb 8\,n - 3\Rb} \,q^{8\,n - 4}
\eea

From \eq{MP13} we can obtain 
\beq \label{MP14}
\Sigma\Lb q\Rb\,\,=\,\,\sum^{\infty}_{n=1}\,\Sigma_n(q)\,\,=\,\,
\,\pi^{5/2}\,\sum^{\infty}_{n=1}\,\Lb \frac {8 \cdot 24}{\pi^3}\,A^2_4\Rb^n\,\frac{\,\Gamma\Lb n+1\Rb\Gamma\Lb 4\,n - 3/2\Rb}{\Gamma\Lb 4\,n \Rb\,\,\Gamma\Lb 8\,n - 3\Rb} \,q^{8\,n - 4}\,
\eeq

The sum of \eq{MP14} converges well since at large $n$ the general term decreases as $n^{-6 n -3/2}$ and leads to a function that increases  with the mass of produced particles $q$. However, this increase cannot continue to large masses due to the unitarity  corrections\cite{VZ} (or due to  the quantum effects \cite{QCINST}). One can see that the instanton--induced vertex $\Gamma(2 \to n)$  has correction of the type shown in \fig{instun} and the
resulting $\Sigma(q)$ cannot be given by \eq{MP14} but instead is equal to
\beq \label{MP15}
\Sigma_R(q)\,\,=\,\,\frac{Im \Sigma(q)}{(1  - I \bigotimes Re \Sigma(q))^2 + (I \bigotimes Im \Sigma(q))^2 }\,\,\xrightarrow{q \to \infty}\,\,0
\eeq
where the imaginary part of $\Sigma(q)$ is given by \eq{MP14} and $\bigotimes$ stands for all needed integrations; we have introduced the notation $I$ for the following integral
\beq \label{I}
I\,\,\,=\,\,\int \frac{d z}{z^5} \,d^4 x\,e^{ - \frac{2 \pi}{g_s}}\,\,\equiv\,\,\int \frac{d z}{z^5} \,d^4 x\,e^{ - \kappa}
\eeq

As will become clear in the next subsection, we actually will need to calculate $\int d q^2 Im \Sigma(q)$. 
The integrand here is expected to rapidly fall off above $q = E_{sph}$ (see \eq{ESPH}) that is the height of the sphaleron barrier setting a typical energy scale for the instanton--induced processes.
It has been shown explicitly \cite{VZ,QCINST} that at $E\,>\,E_{sph}$  $\Sigma(q)$ indeed starts to fall off.

Finally we obtain
\beq \label{MP16}
\int^{E^2_{sph}}\,d q^2\,\Sigma(q)\,\,\,=\,\,\,\pi^{5/2}\,\sum^{\infty}_{n=1}\,\Lb \frac{8 \cdot 24}{\pi^3}\,A^2_4\Rb^n\,\frac{\Gamma\Lb n+1\Rb\Gamma\Lb 4\,n - 3/2\Rb}{\Gamma\Lb 4\,n  \Rb\,\,(4n - 1)\,\Gamma\Lb 8\,n - 3\Rb} \,E^{8\,n - 2}_{sph}\,
\eeq

\FIGURE[h]{\begin{minipage}{140mm}
\centerline{\epsfig{file=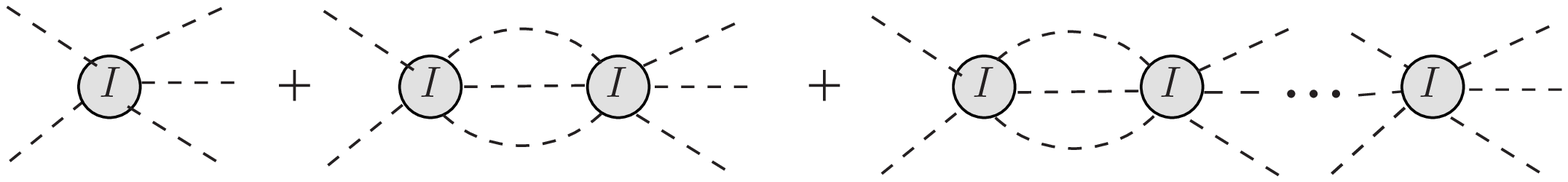,width=140mm}}
\caption{The unitarity correction to the instanton induced vertex $\Gamma(2 \to n)$. }
\end{minipage}
\label{instun}}

Using \eq{ESPH} and the exprerssion for $A_4$ from \eq{MP7} we can rewrite \eq{MP16} in more convenient for for further calculations, namely
\beq \label{MP17}
\int^{E^2_{sph}}\,d q^2\,\Sigma(q)\,\,\,=\,\,\,\pi^{5/2}\,z_k\,\bar{z}_k\,\sum^{\infty}_{n=1}\,
\frac{\Gamma\Lb n+1\Rb\Gamma\Lb 4\,n - 3/2\Rb}{\Gamma\Lb 4\,n \Rb\,\,(4n - 1)\,\,\Gamma\Lb 8\,n - 3\Rb}\,\,
\Lb \frac{1}{48\,\pi^3}\Rb^n\,\kappa^{10\,n - 2}
\eeq
\subsection{Multiperipheral kinematics and the interaction between instantons}

\FIGURE[h]{\begin{minipage}{70mm}
\centerline{\epsfig{file=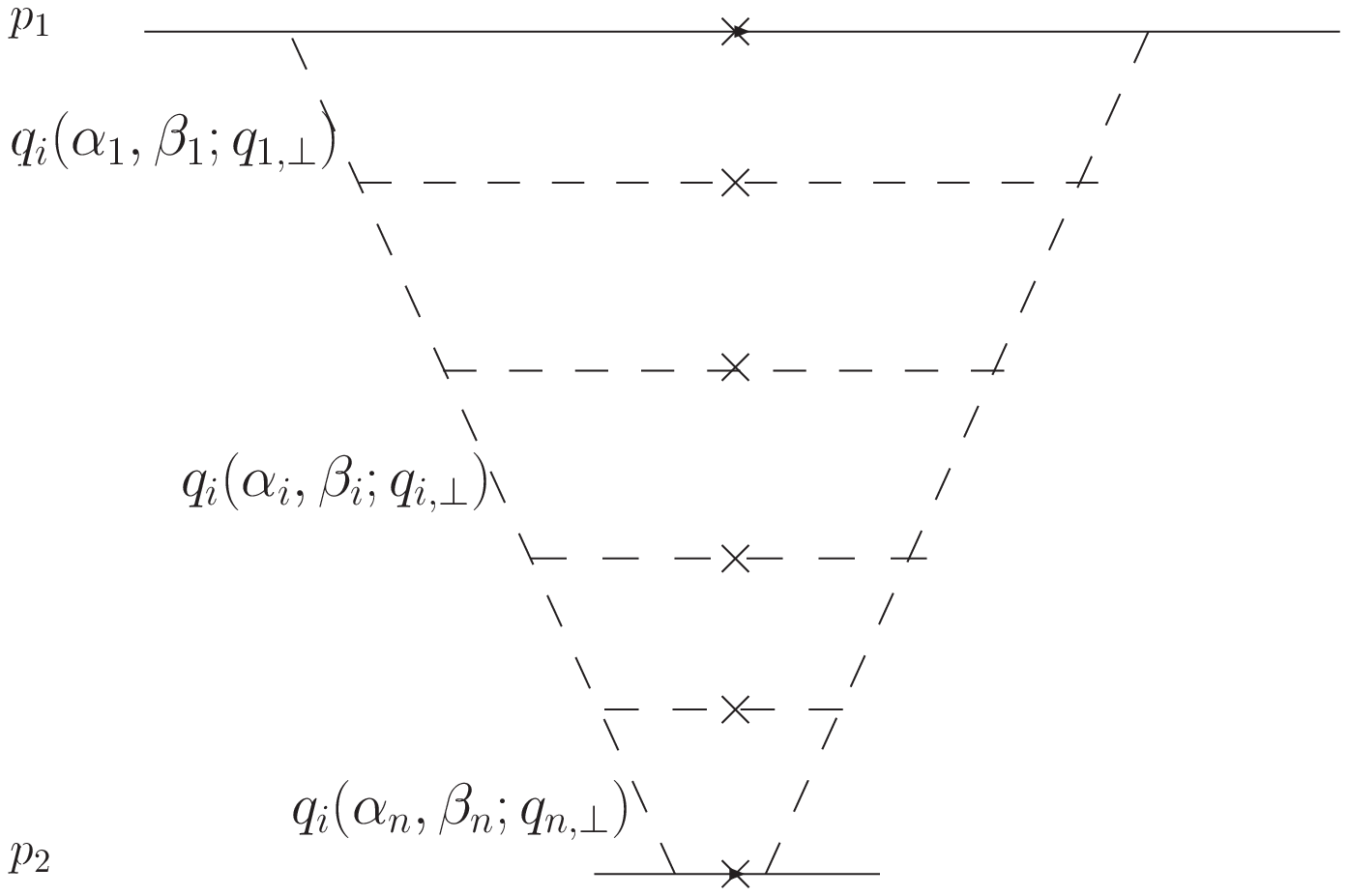,width=60mm}}
\caption{ The simplest multiperipheral diagram for scalar particle interaction. Crosses indicate that the particle is on mass shell.}
\end{minipage}
\label{instmpd}}

In this section we recall the calculation of the simple multiperipheral ladder diagram of \fig{instmpd}  for scalar particles in the momentum representation. Since we are interested in the total cross section the measure of integration over produced particles with momenta $k_i = q_{i -1} - q_{i}$ has the form  $\delta(k^2 - m^2)\,d^4 k/(2\pi)^4$. For momenta $q_i$ we use the Sudakov decomposition, namely,
\beq \label{MPD1}
q_{i,\mu}\,\,=\,\,\alpha_i \, p_{1,\mu}\,\,+\,\,\beta_i\,p_{2,\mu}\,\,+\,\,q_{i,\perp;\mu}
\eeq
It turns out that the maximal contribution  for the diagrams of \fig{instmpd}-type stems from the following region of integration with respect to $\alpha_i$ and $\beta_i$:
\beq \label{MPD2}
\alpha_1\,\gg\,\alpha_2\,\gg \,\dots\, \gg\,\alpha_i\,\gg\,\dots\,\gg\,\alpha_n;\,\,\,\,\,\,\,\,\,\,\,\,\,\,\,\,
\beta_1\,\ll\,\beta_2\,\ll \,\dots\, \ll\,\beta_i\,\ll\,\dots\,\ll\,\beta_n;
\eeq
One can see that the integration can be performed using momenta $q_i$ with the phase space $d \alpha_i \, d \beta_i s\,d^2 q_{i,\perp}/(2 \pi)^4$  where $s = (p_1 + p_2)^2$ and 
\bea \label{MPD3}
q^2_i\,\,&=&\,\,\alpha_i \beta_i\,s\,+\,q^2_{i, \perp}\,\,=\,\,q^2_{i, \perp}\,\,+\,\,{\cal O}\Lb m^2 \frac{\alpha_{i }}{\alpha_{i+1}} \Rb\,; \\
k^2_i \,\,&=&\,\,(\alpha_i\,\beta_{i+1}\,+\,\,\alpha_{i + 1}\,\beta_{i})\,s - k^2_{i,\perp}\,\,=\,\,\alpha_i\,\beta_{i+1}\,s \,-\,k^2_{i,\perp}\,\,+\,\,{\cal O}\Lb  m^2 \frac{\alpha_{i }}{\alpha_{i+1}}\Rb \nonumber
\eea
Therefore, the diagram of \fig{instmpd} leads to the following expression
\bea \label{MPD4}
&&A^2_n\Lb \fig{instmpd}\Rb\,\,= \nonumber\\
&&\,\,\,\, =\,\,g^4\,\prod^n_{i=1}\,\int \,d \alpha_i\,d\,\beta_i s\,\,\frac{ d^2 q_{i,\perp}}{(2 \pi)^3}
\delta\Lb k^2_i - m^2\Rb \,\,\,\frac{g^2}{(q^2_i - m^2)^2}\,\,g^2 \delta\Lb p_1 - q_1)^2 - m^2 \Rb\,\g^2\delta \Lb p_2 + q_n)^2 - m^2\Rb\nonumber \\
 &&\xrightarrow{s \gg m^2}  \,\,g^4
\prod^n_{i=1}\,\int \,\frac{d \alpha_i\,}{\alpha_i} \frac{d^2 q_{i,\perp}}{(2 \pi)^3}
\,\,\frac{g^2}{(q^2_{i,\perp}+ m^2)^2}\,\,g^2 \delta\Lb\alpha_1\,s - m^2 - q^2_{1,\perp}\Rb\,\delta\Lb\beta_n\,s  - m^2 - q^2_{n,\perp}\Rb\nonumber \\
&& \,\,\,\, =\,\,\frac{1}{n!}\,\frac{g^4}{s}\Lb \int^1_{\alpha_n = m^2/s} \frac{d \alpha_i\,}{\alpha_i} \frac{d^2 q_{i,\perp}}{(2 \pi)^3}\frac{d \alpha_i\,}{\alpha_i} \,\,\frac{g^2}{(q^2_{i,\perp}+ m^2)^2}\Rb^n\,\nonumber\\
&& \,\,\,\, =\,\,\,\frac{g^4}{s}\frac{\ln^n(s/m^2)}{n!}\,\Lb  \frac{d^2 q_{i,\perp}}{(2 \pi)^3}\frac{g^2}{(q^2_{i,\perp}+ m^2)^2}\Rb^n
\eea

Summing over $n$ we get the total cross section:
\beq \label{MPD5}
\sigma_{tot}\,\,=\,\,\frac{1}{s}\,\sum^{\infty}_{n=1}\,A^2_n\,\,=\,\,g^4\, \Lb\frac{s}{m^2}\Rb^{\Delta - 2},
\eeq
where
\beq \label{MPD51}
\Delta \,\,\,=\,\,\,\int\, \frac{d^2 q_{i,\perp}}{(2 \pi)^3}\frac{g^2}{(q^2_{i,\perp}+ m^2)^2}.
\eeq
The expression for $\Delta$ can be re-written in the impact parameter ($b$)  representation:  replacing the particle propagator in $q$ -representation ($G(q)=1/(q^2_{\perp} + m^2) $) by the propagator 
\beq \label{B}
G( q)\,\,\xrightarrow{ q \to b} \,\,G\Lb b \Rb \,\,\,=\,\,\,\int \,\frac{d^2 q_{\perp}}{(2 \pi)^2}\,e^{ i \vec{q}_\perp \cdot \vec{b}}\,\frac{ 1}{q^2_{\perp} + m^2}\,\,=\,\,\frac{1}{2 \pi}\,K_0\Lb m\,b\Rb
\eeq
we obtain that
\beq \label{MPD6}
\Delta \,\,\,=\,\,\,\int\, \frac{d^2 q_{i,\perp}}{(2 \pi)^3}\frac{g^2}{(q^2_{i,\perp}+ m^2)^2}\,\,=\,\,g^2\,\int d^2 b\,\, G^2\Lb b\Rb
\eeq
 The diagram of \fig{ladcr} is more complicated. First, we do not produce one particle but the bunch of particles that we represent by an unparticle. However using the dispersion relation for $\Sigma(q)$ we can replace 
\beq \label{MPD7}
\frac{1}{m^2 - k^2 }\,\,\longrightarrow\,\,\frac{1}{2\pi}\,\int \frac{Im \,\Sigma(M)\,\,d M^2}{ M^2 -  k^2};\,\,\,\,
\delta\Lb m^2 - k^2\Rb \,\,\longrightarrow\,\,\int \,d M^2\delta\Lb M^2 - k^2\Rb\,\,Im \,\Sigma(M)
\eeq
Integrating over $\beta_i$ we obtain an extra factor $\int \,d M^2\,\,Im\, \Sigma(M)$.

A natural question arises about what should we use instead of $G(b)$ in \eq{MPD6}. As has been discussed above, the instanton field is equal to \cite{INSTD}

\beq \label{MPD8}
\phi_{inst}(z_k,x_k; z, x)\,\,=\,\,\frac{2 \pi}{g_s}\,G\Lb z_k,x_k; z, x\Rb
\eeq
where $G\Lb z_k,x_k; z, x\Rb$ is the dilaton propagator in $AdS_5\times S^5$ theory. Therefore to find the Green's function for the interaction of two instantons we need to evaluate the following product:
\bea \label{MPD9}
&&\phi_{inst}(z_k,x_k; z, x)\,G^{-1}\Lb z_k,x_k; z, x\Rb \,\times\, G\Lb z, x; z', x' \Rb\,\times\, G^{-1}\Lb z', x'; z_{k-1},x_{k-1} \Rb\,\phi_{inst}(z_{k-1},x_{k-1}; z', x')\,\,=\nonumber\\
&&\,\,\Lb \frac{2 \pi}{g_s}\Rb^2\,G\Lb z_k, x_k; z_{k-1}, x_{k-1} \Rb\,\equiv\,\,\kappa^2\,G\Lb z_k, x_k; z_{k-1}, x_{k-1} \Rb
\eea
It is clear that this procedure is just the same as we did in \eq{MP5}-\eq{MP7} for unparticles that have been produced on the boundary.

As we have learned the most convenient way to factor out the energy dependence is to perform all our calculation in the mixed representation: two energy variables $(\alpha_i$ and $\beta_i$ (or $q_{i,+}$ and $q_{i,-}$)  and  coordinates. In $AdS_5$ such propagator has been found (see Refs.\cite{BST2,BST3,LMKS}):
\beq \label{MPD10}
G\Lb z_k, x_k; z_{k-1}, x_{k-1} \Rb\,\,=\,\,G_{3}\Lb u \Rb \,\,=\,\,\frac{1}{4 \pi}\,\frac{1}{\left\{ 1 + u + \sqrt{u (u + 2)}\right\}^2\,\sqrt{u (u + 2)}}
\eeq
with

\beq \label{UTR}
u\,\,=\,\,\frac{ (z_k - z_{k-1})^2 + b^2}{ 2 \,z_{k}\,z_{k -1}}
\eeq
where $b = x_{k-1,\perp}\,\,-\,\,x_{k,\perp}$ is the impact parameter  which is a  variable conjugated to $q_i$ in \fig{instmpd}. 

For the vertex $\Gamma(2 \to n)$ in \fig{ladcr} we need to collect factors from \eq{MPD9} and \eq{I}. Doing so we obtain that we need to replace $g$ in \eq{MPD4} by
\beq  \label{MPD11}
g \,\,\,\longrightarrow\,\,\,\kappa^2\, \int \frac{d z_k}{z^5_k}
\eeq

Concluding this section we can claim that $|A|^2$ for the amplitude of \eq{MP2} can be written in the multiperipheral form of \eq{MPD3} and \eq{MPD4} with the following short glossary
\bea \label{GLO}
\int \,\delta(m^2 - k^2)  & \,\,\,\to \,\,\,& \int \,d q^2\,\, \Sigma(q) \,\,\,\Lb\mbox{see \eq{MP17}}\Rb\,;\nonumber \\
G(b)\,\, \Lb \mbox{see \eq{B}}\Rb & \to & G(u) \, \,\,\Lb\mbox{see \eq{MPD10}} \Rb\,;\nonumber \\
g\,\,\,\,\,\,\,\, &\to &\kappa^2\, \int \frac{d z_k}{z^5_k} \, \,\,\Lb \mbox{see \eq{MPD11}} \Rb
\eea
\newpage 
\section{The energy dependence of the multiparticle production cross section }

\subsection{Equation and its solution}

~

\FIGURE[h]{\begin{minipage}{55mm}
\centerline{\epsfig{file=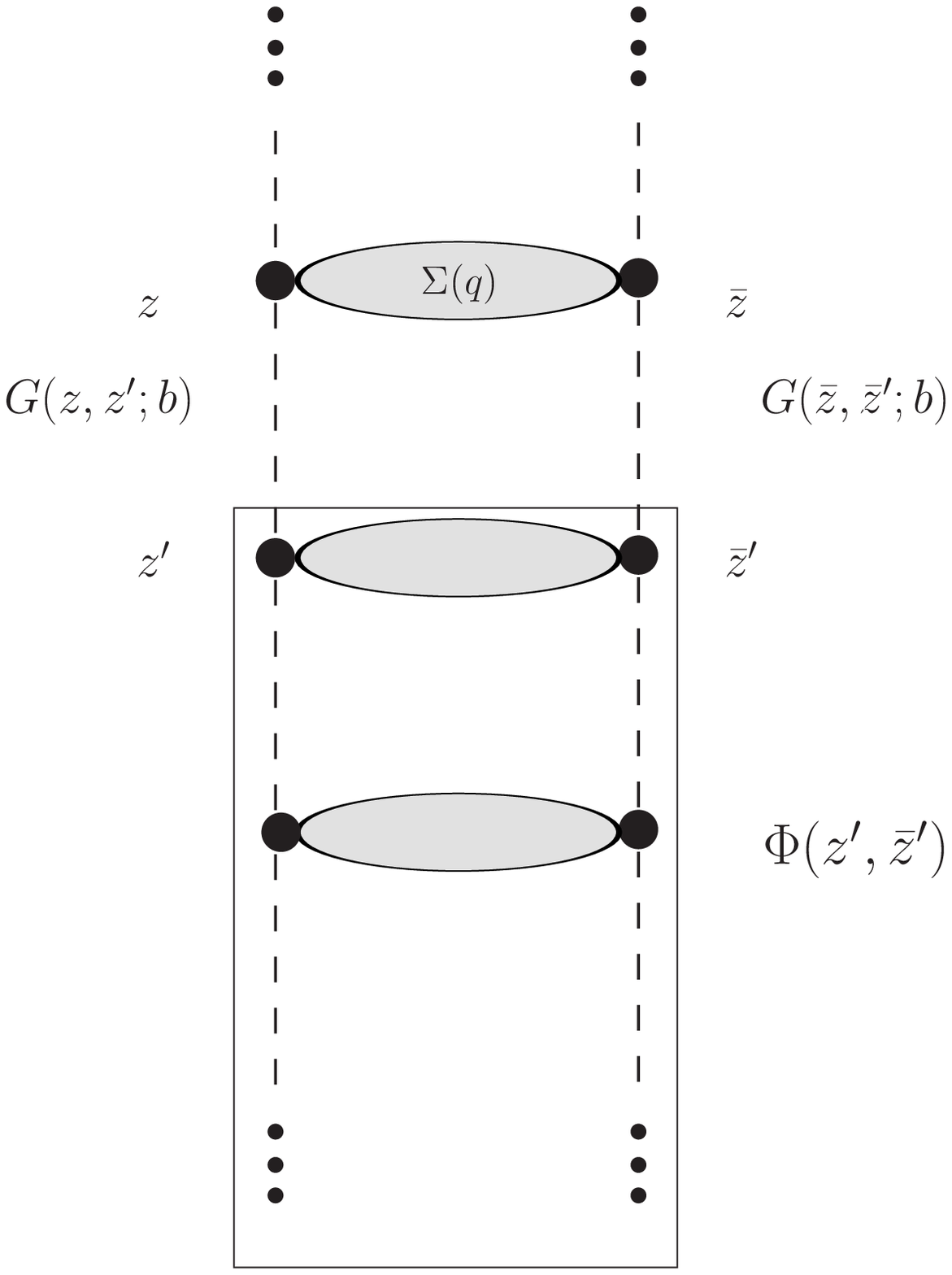,width=50mm, height=50mm}}
\caption{ The equation for the scattering amplitude. The black crosses denote the instanton vertices. $\Sigma(q)$
is given by \protect\eq{MP14} while propagators $G$ are written in \protect\eq{MPD10}.\label{insteq}}
\end{minipage}
}

For the imaginary part of the scattering amplitude we can write down the equation which is illustrated  in \fig{insteq}.
It has a simple form, namely
\beq \label{SMP1}
\frac{d\,\Phi\Lb z,\bar{z}\Rb}{ d \ln s}\,\,=\,\,\Lb\frac{2 \pi}{g_s}\Rb^4\,e^{- 2\,\frac{2 \pi}{g_s}}\int\,\frac{d z'}{z'^5}\,\frac{d\bar{z}'}{\bar{z}'^5}\,\,K\Lb z,\bar{z}; z', \bar{z}'\Rb\,\, 
\Phi\Lb z',\bar{z}'\Rb
\eeq
where  the kernel $K$ is given by
\beq \label{SMP2}
K\Lb z,\bar{z}; z', \bar{z}'\Rb\,\,=\,\,\int\,d^2 b\,G\Lb z,z'; b \Rb\,G\Lb \bar{z},\bar{z}'; b \Rb\,\,\int d q^2\,\, \Sigma(q)\,\
\eeq
As we have seen in \eq{MP17} $ \int d q^2\, \Sigma(q)$ is proportional to  $z' \bar{z}'$. Introducing $\int \,d q^2\,\Sigma(q)=
z' \bar{z}'\, \tilde{\Sigma}(\kappa)$ we can rewrite \eq{SMP1} in the form
\beq \label{SMP3}
\frac{d\,\Phi\Lb z,\bar{z}\Rb}{ d \ln s}\,\,=\,\,\kappa^4\,e^{- 2 \kappa}\,\tilde{\Sigma}(\kappa)\int\,\frac{d z'}{z'^4}\,\frac{d\bar{z}'}{\bar{z}'^4}\,\,\tilde{K}\Lb z,\bar{z}; z', \bar{z}'\Rb\,\, 
\Phi\Lb z',\bar{z}'\Rb
\eeq
with
\beq \label{SMP4}
\tilde{K}\Lb z,\bar{z}; z', \bar{z}'\Rb\,\,=\,\,\int\,d^2 b\,G\Lb z,z'; b \Rb\,G\Lb \bar{z},\bar{z}'; b \Rb
\eeq

~

The general form of this kernel is rather complicated but in the region where $z \gg z'$ and $\bar{z} \gg \bar{z}'$ the integral over $b$ can be taken explicitly and it leads to
\beq \label{SMP5}
\tilde{K}\Lb z,\bar{z}; z', \bar{z}'\Rb\,\,=\,\,\frac{1}{2 \pi}\,\frac{t^3 t'^3}{t^5}\,F\Lb x\,=\,\frac{z}{\bar{z}}\Rb 
\eeq
where
\beq \label{SMP6}
F\Lb x\,=\,\frac{z}{\bar{z}}\Rb \,\,=\,\, \frac{ -1 + 8 x^2 - 8 x^6 + x^8 +12 x^4\ln(x^2)}{4 ( x^2 - 1)^5}
\eeq
with $t = z\,\bar{z}$ and $t' = z'\,\bar{z'}$.
Therefore, to study the main features of the kernel $\tilde{K}$ of \eq{SMP4} we can use the simplified formula
\beq \label{SMP7}
\tilde{K}\Lb z,\bar{z}; z', \bar{z}'\Rb\,\,=\,\,\frac{1}{2 \pi}\,\left\{ \frac{t'^3}{t^2}\,F(x)\,\Theta\Lb t - t'\Rb\,\,+\,\,
\frac{t^3}{t'^2}\,F(x')\,\Theta\Lb t' - t\Rb \right\}
\eeq 
where $x' = z'/\bar{z}' $.

\FIGURE[h]{\begin{minipage}{55mm}
\centerline{\epsfig{file=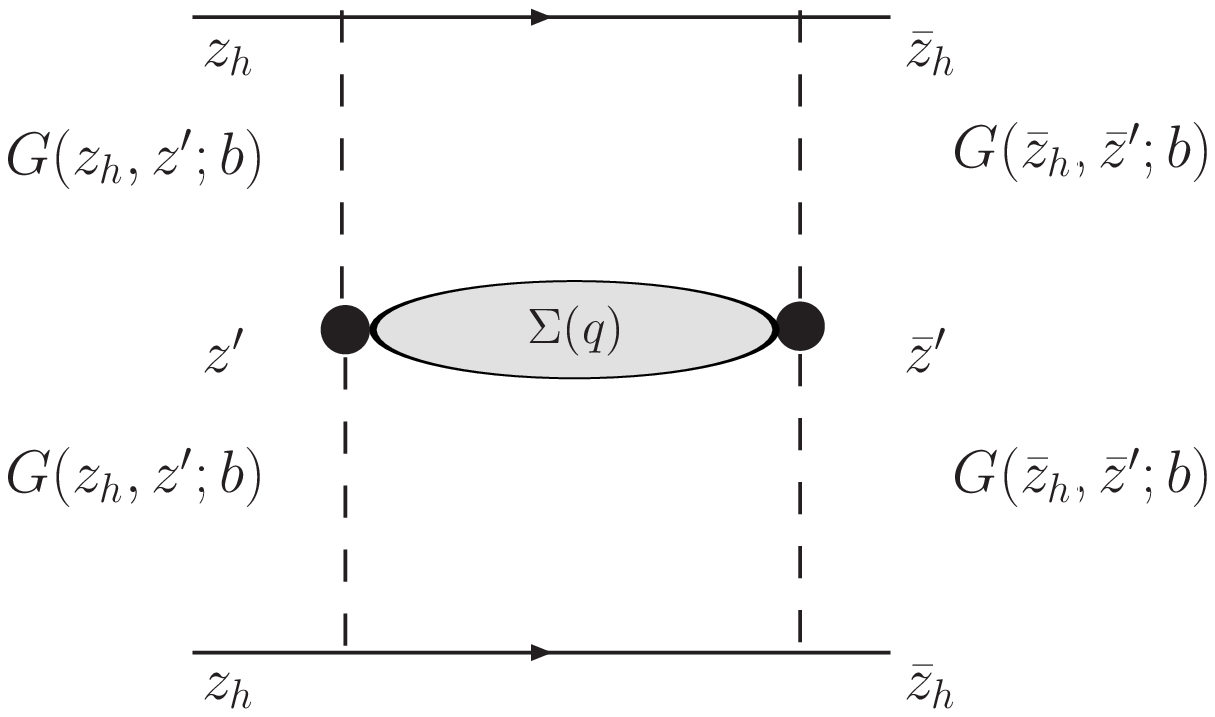,width=50mm,height=40mm}}
\caption{ The  simplest diagram of the instanton contribution (the first interation of \protect\eq{SMP1}). The black crosses denote the instanton vertices. $\Sigma(q)$
is given by \protect\eq{MP14} while propagators $G$ are written in \protect\eq{MPD10}.\label{insteqfd}}
\end{minipage}
}
We  start the analysis of the solution to \eq{SMP1} by calculating the first iteration of this equation given by \fig{insteqfd}.
For the region of integration over $t' < t_h$ we have the contribution

\bea\label{SMP8}
& & A^2\Lb \fig{insteqfd}; t_h , x_h\Rb\,\,= \nonumber\\
\,\,\,\,\,\,\,\,\,\,& =&\,\,\,\,\,\,\frac{\ln(s/s_0)}{2\pi\,s}\,\,\kappa^4\,e^{- 2 \kappa}\,\tilde{\Sigma}\,(\kappa)\,\int^{z_h} d z' \int^{\bar{z}_h} d \bar{z}' \,\frac{1}{t'^4}\,
F^2(x_h)\frac{t^6_h \,t'^6}{t^{10}_h}\,\nonumber\\
\,&=&\,\,\frac{\ln(s/s_0)}{2 \pi\,s}\,\,\,\,\kappa^4\,e^{- 2 \kappa}\,\tilde{\Sigma}\,(\kappa)\,\frac{1}{3}\,F^2(x_h)\frac{1}{t_h}
\eea

In the case when  $t' >  t_h$ one can see that
\bea \label{SMP9}
 & & A^2\Lb \fig{insteqfd}; t_h , x_h\Rb\,\, =\nonumber\\
 \,\,\,\,\,\,\,\,\,\,\,\,& = &\frac{\ln(s/s_0)}{2 \pi \,s}\,\,\kappa^4\,e^{- 2 \kappa}\,\tilde{\Sigma}\,(\kappa)\,\ \,\int_{z_h} d z' \int_{\bar{z}_h} d \bar{z}' \,\frac{1}{t'^4}\,
F^2(x')\frac{t^6_h \,t'^6}{t'^{10}_h}\nonumber\\
 \,\,&=&\,\,\frac{\ln(s/s_0)}{2 \pi\,s}\,\,\kappa^4\,e^{- 2 \kappa}\,\tilde{\Sigma}\,(\kappa)\,\frac{1}{7}\,F^2(x_h)\frac{1}{t_h}
\eea

\begin{flushleft}
The lesson from these simple estimates is clear: the typical $z'$ and $\bar{z}'$ are about $z_h$ and  $\bar{z}_h$,
respectively. Actually, we need to integrate over $z_h$ and $\bar{z}_h$ separately for target and projectile and 
for high energy scattering $z_h \simeq \bar{z}_h$ (see Refs.\cite{BST2,LMKS}).  However for our purposes the simple estimates given above suffice.
\end{flushleft}
Comparing the sum of \eq{SMP8} and \eq{SMP9} with the diagram without the instanton which in our notation is
\beq \label{SMP10}
A^2(\mbox{without instanton})\,\,=\,\,t_h\,F(x)/s
\eeq
we can conclude that
\bea \label{SMP11}
&&A^2\Lb \fig{insteqfd}\Rb \,\,=\,\,A^2(\mbox{without instanton})\,\,\ln(s/s_0)\,\Lb \kappa^4\,e^{- 2 \kappa}\tilde{\Sigma}(\kappa)\,\frac{10}{21}\,\frac{F(x_h)}{t^2_h}\Rb\,\nonumber \\
&&=\,\,A^2(\mbox{without instanton})\,\,\Delta\,\ln(s/s_0)
\eea
with 
\beq \label{SMP12}
\Delta \,\,=\,\, \frac{1}{2 \pi}\kappa^4\,e^{-2 \kappa}\,\tilde{\Sigma}(\kappa)\,\frac{10}{21}\,\frac{F(x_h)}{t^2_h}
\eeq
 In this estimate we assumed that the final answer will be $A^2 \propto s^{\Delta -1}$.  In spite of the fact that these simple estimates are  very instructive we should not expect them to give a reliable quantitative result since we made a strong assumption that $z$ and $\bar{z}$ in the target and  projectile are the same.

For the solution of \eq{SMP1} we assume that 
\beq \label{SMP13}
\int\,d\,r\,\Phi\Lb z,\bar{z}\Rb\,\,=\,\,z^2 \,\bar{z}^2 \tilde{\Phi}\Lb z\,\bar{z};s \Rb
\eeq
where $r = \bar{z}/z$,
and looking for 
\beq \label{SMP14}
 \tilde{\Phi}\Lb t=z\,\bar{z}\Rb\,\,=\,\,\phi(t_h=z_h  \bar{z}_h)\,s^{\Delta}
\eeq

\FIGURE[t]{
\centerline{\epsfig{file=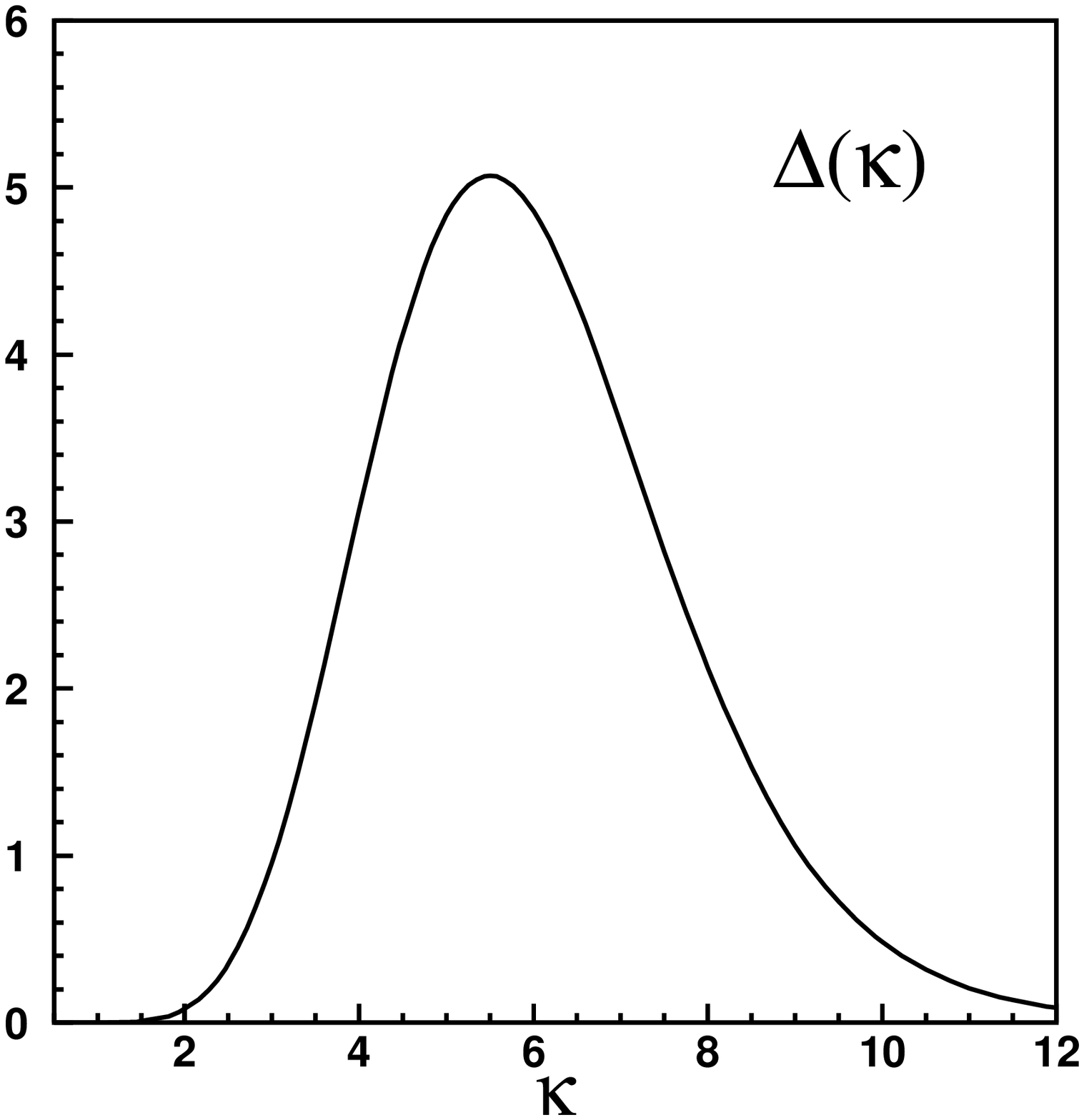,width=100mm,height=70mm}}
\caption{ $\Delta(\kappa)$ versus $\kappa$, for $z^4_h = 2\, \alpha'^2$ and $N_c =3$.}
\label{delta}}

For $\tilde{\phi}\Lb t \Rb$  \eq{SMP1} has the form
\beq \label{SMP15}
t^2\,\frac{d\,\tilde{\phi}\Lb t; s \Rb}{ d \ln s}\,\,=\,\,\Lb\frac{2 \pi}{g_s}\Rb^4\,e^{- 2\,\frac{2 \pi}{g_s}}\int\,d t'\,K\Lb t'\Rb\,\, 
\tilde{\phi}\Lb t' ; s \Rb\,\,
\eeq
where
\beq \label{SMP16}
K \Lb t \Rb\,\,=\,\,
\,\,\int\,d r\frac{ d z'}{z'^2}\,\frac{d \bar{z}'}{\bar{z}'^2}\,\,d^2 b\,\,G\Lb z,z'; b \Rb\,G\Lb \bar{z},\bar{z}'; b \Rb\,\,\tilde{\Sigma}(q)\,\
\eeq
\eq{SMP15} stems from \eq{SMP1} after integration over $r$ of both parts of this equation.

The integrals over $z'$ and $\bar{z}'$ can be taken by switching to a new variable
$y =  2 z z'\,\Lb  u + 1 + \sqrt{u(u + 2)}\Rb$. With this variable the integrals can be easily evaluated:
\beq \label{SMP17}
 I\Lb z; b \Rb\,\equiv\,\,\int\,\frac{ d z'}{z'^2}\,\,\,G\Lb z,z'; b \Rb\,=\,\frac{4 z^2}{4 \pi}\,\int^{\infty}_{2(z^2 + b^2)}\,\,\frac{d y}{ y^2\,(y - 2 \,z^2)}\,\,=\,\,\frac{1}{4 \pi}\
\frac{z^2 +(z^2 + b^2)\,\ln\frac{b^2}{z^2 + b^2}}{z^2\,(z^2 + b^2)}
\eeq
and 
\beq \label{SMP18}
K \Lb t \Rb\,\,=\,\,\int\,d r\,\int d^2 b \, I\Lb z; b \Rb \,I\Lb \bar{z}; b \Rb\,\,=\,\,\frac{1}{16 \pi}\,\int\,d r\,\frac{1}{r ( 1 - r^2)}\,\ln\Lb 1/r\Rb\,\,=\,\,\frac{1}{16 \pi}\,\frac{\pi^2}{12}\,=\,\,\frac{\pi}{16\,12}
\eeq

Finally,   we obtain for 
$ \Delta$ the following equation
\beq \label{SMP19}
\Delta\,\,=\,\,\,\frac{\pi}{16\,12}\,\frac{1}{t^2_h}\,\kappa^4\,e^{- 2 \kappa}\,\tilde{\Sigma}(\kappa)
\eeq

\subsection{The Pomeron intercept}
\eq{SMP15} gives us the energy dependence of the cross section, namely,
\beq \label{PI1}
\sigma_{inelastic}\,\,\propto\,\,s^{ \Delta -2}
\eeq
Therefore our mechanism of multiparticle production can be important at high energies only if $\Delta $ exceeds 2. At first sight $\Delta$ has a dimension since it is proportional to $ 1/z^4_h$. This
is the artifact of our simplification since we put $L =1$. Restoring $L$ explicitly, we get the following expression for $\Delta$:
\beq\label{PI2}
\Delta\,\,\,=\,\,\,\frac{\pi}{16\,12}\,\frac{L^4}{t^2_h}\,\kappa^4\,e^{- 2 \kappa}\,\tilde{\Sigma}\Lb\kappa\Rb\,\,=\,\,\frac{\pi}{16\,12}\,\lambda\frac{\alpha'^2}{t^2_h}\,\kappa^4\,e^{- 2 \kappa}\,\tilde{\Sigma}(\kappa)\,
 \,\,=\,\,\frac{\pi^3}{24}\, \,N_c\,\frac{\alpha'^2}{t^2_h}\,\kappa^3\,e^{- 2 \kappa}\,\tilde{\Sigma}(\kappa)
\eeq

To take account of the instanton interactions with axions, we need to multiply \eq{PI2} by factor of 4  ( see \eq{SG1}). Indeed, we have four transitions in the $t$-channel:  $2 d \to 2 d$, $2 d \to 2 a$,$2 a \to 2 a$ and $2 a \to 2 d$ ($d$ and $a$ denote the dilaton and axion respectively).

In \eq{PI2} we used  that $\lambda \,=\,g^2_{YM}\,N_c\,=\,\alpha_{YM}\,4 \pi\,N_c\,=g_s\,4 \pi N_c\,\,=\,\,8 \pi^2 N_c/\kappa$ and $L^4 \,=\,\alpha'^2 \,\lambda$, where $\alpha'$ is the slope of the Reggeon trajectory, $g_s$ is a string coupling, and $g_{YM}$dinstmp.tar is the YM coupling.

In \fig{delta} we plot $\Delta(\kappa)$ assuming that $z^2_h =0.5\ \alpha' $ , and $N_c = 3$. We believe that this case can resemble the real world (see discussion below). One can see that for $\kappa > 4$  $\Delta(\kappa) > 2$ and therefore the instanton--induced multiparticle production has a cross section that increases with energy.

At $\kappa =2$ $\lambda = 120$ but the value of $g_s \approx 3$ is rather large for the approach that can be reduced to weak gravity. We recall that this approach corresponds $g_s <1$ but $g_s N_c >1$.  From \fig{delta} one can see that at $\kappa = 8$ we have $g_s \approx 0.75 $ while $g_s N_c = 2.25 >1$. For this value of $g_s$ $\lambda \approx 30$ and the graviton interaction generates $(2 /\sqrt{\lambda}) \approx 15 - 20\%$ of the cross section.  Our approach can be trusted if  $g_s <1$ but $g_s N_c >1$; this translates into the window for $\kappa$ in \fig{delta} is $6\,<\,\kappa\,<\,18$.  One can see that for values $6\,<\,\kappa\,<\,8$  which are in this window $\Delta - 2\,\,> \,0$ leading to the cross section for multiparticle production that increases with  energy.

These estimates show that we can obtain phenomenologically reasonable values of $\Delta$. However, these values  crucially depend on the value for $z_h$ which gives us a new scale for high energy interaction.   We cannot calculate the wave functions of the observed hadrons in N=4 SYM (remember that this theory does not possess confinement). However in our estimates we implicitly assumed that the hadron wave functions are peaked at $z=z_h$.                                              
As has been seen from our calculation of the Pomeron intercept and as we will demonstrate in  the next section calculating the slope of the new Pomeron trajectory, $z_h$ is our key phenomenological parameter that determines all features of the Pomeron. We choose $z^2_h = 0.5\, \alpha'$ since it gives
reasonable Pomeron intercept and $\alpha'_P \,\approx 0.5\, \alpha'$ as it should be for the closed string in the string theory that corresponds to N=4 SYM in the limit $\lambda \to \infty$.

\subsection{The Pomeron slope}
\eq{SMP1} allows us to find the slope of ``Pomeron" trajectory $\Delta(t) = \Delta + \alpha'_P\,t$ where $t = - Q^2$ is the momentum transferred along the Pomeron. However we need to rewrite the equation for $t \neq 0$; this leads to
\beq \label{PSL1}
\frac{d\,\Phi\Lb z,\bar{z}\Rb}{ d \ln s}\,\,=\,\,\Lb\frac{2 \pi}{g_s}\Rb^4\,e^{- 2\,\frac{2 \pi}{g_s}}\int\,\frac{d z'}{z'^5}\,\frac{d\bar{z}'}{\bar{z}'^5}\,\,K\Lb z,\bar{z}; z', \bar{z}' | Q\Rb\,\, 
\Phi\Lb z',\bar{z}'\Rb
\eeq
with
\beq \label{PSL2}
K\Lb z,\bar{z}; z', \bar{z}'|Q\Rb\,\,=\,\,\int\,d^2 b\,\,e^{i \vec{Q} \cdot\vec{b}}\,\,G\Lb z,z'; b \Rb\,G\Lb \bar{z},\bar{z}'; b \Rb\,\,\int d q^2\,\, \Sigma(q)\,\
\eeq
Actually \, for calculation of the slope we need only to find the second term in the expansion of $K\Lb z,\bar{z}; z', \bar{z}'|Q\Rb$  with respect to $Q$, namely,
\bea \label{PSL3}
&&K\Lb z,\bar{z}; z', \bar{z}'|Q\Rb\,\,= \\
&&\,\,\,\,\,=\,\,\int\,d^2 b\,\,G\Lb z,z'; b \Rb\,G\Lb \bar{z},\bar{z}'; b \Rb\,\,\int d q^2\,\, \Sigma(q)\,\,+\,\,\frac{1}{4}\,Q^2\,\int\,b^2\,d^2 b\,\,G\Lb z,z'; b \Rb\,G\Lb \bar{z},\bar{z}'; b \Rb\,\,\int d q^2\,\, \Sigma(q)\,\nonumber\\
&&\,\,\,\,\,=\,\,K_1\Lb z,\bar{z}; z', \bar{z}'\Rb \,+\,Q^2\,K_2\Lb z,\bar{z}; z', \bar{z}'\Rb\nonumber
\eea
$K_1$ is the kernel of \eq{SMP1} for which we have the explicit form given by \eq{SMP5} for $z\, \gg\, z'$ and $\bar{z}\,\gg\,\bar{z}'$. In the same kinematic region $K_2$ has the form
\beq \label{PSL4}
K_2 \Lb t \Rb\,\,=\,\,\frac{1}{2}\int\,d r\,\int\,b^2\, d^2 b \, I\Lb z; b \Rb \,I\Lb \bar{z}; b \Rb\,\,
\eeq

Using \eq{PSL4} we obtain the expression for the slope of the Pomeron trajectory:

\beq \label{PSL7}\,
\alpha'_P\,\,=\,\Lb K_2 \Lb t_h \Rb/K _1\Lb t_h \Rb\,\Rb\,\Delta\,=\,\,0.23\, \ t_h \,\Delta
\eeq
The numerical coefficient is the result of the  numerical evaluation of the integral of \eq{PSL4}. Using our value for $t_h = 0.5 \alpha'$ and $\Delta \approx 4$ which we have for $N_c=3$,  we obtain $
\alpha'_P\,\,\approx 0.5\,GeV^{-2}$.

\subsection{Corrections}

\DOUBLEFIGURE[h]{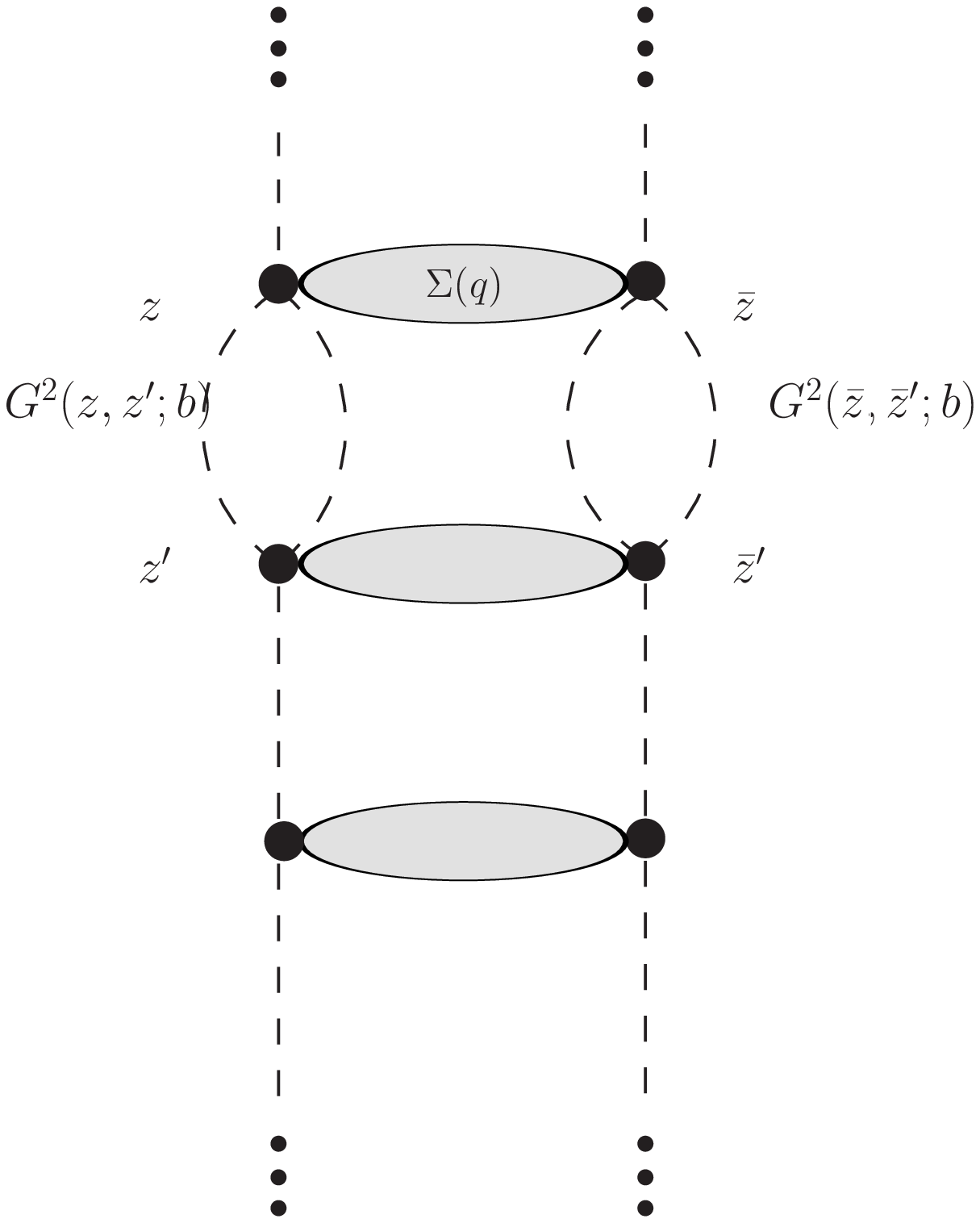,width=50mm,height=40mm}{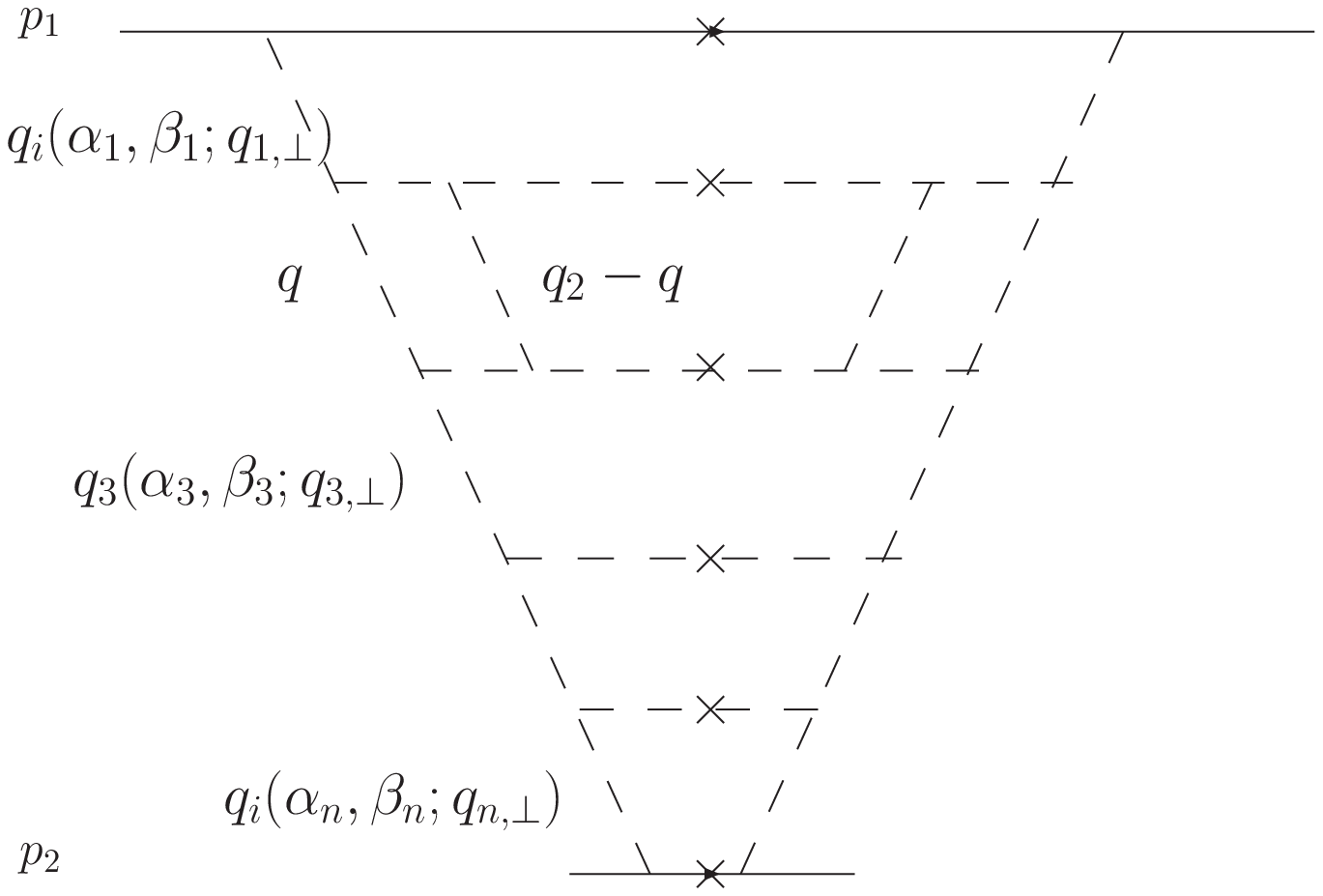,height=40mm}
{ The correction to the amplitude due to exchange of two dilatons. All notation as in \protect\fig{insteq}.\label{instcore}}{The correction to the amplitude in the simple multiperipheral diagram.\label{instcorempd}}

In this subsection we will discuss the possible corrections to the simple multiperipheral diagram of \fig{insteq}.
The main corrections stem from the two--dilaton exchange (see \fig{instcore}). It easy to see that these corrections 
change the value of $\Delta$.

 Indeed, let us consider the simple multiperipheral diagram of \fig{instcorempd}. The extra dilaton in $t$ channel will generate the imaginary amplitude and therefore only the diagram with the exchange of two extra dilatons can contribute to the total cross section. To evaluate it we need to perform the extra integration over $d^4 q = d \alpha_q \,d \beta_q \,s\,d^2 q_{\perp}$ where
\beq \label{COR1}
q_\mu\,\,=\,\,\alpha_q\,p_{1,\mu}\,\,+\,\,\beta_q\,p_{2,\mu}\,\,+\,\,q_{\perp,\mu}
\eeq
We close the  contour of integration over $\alpha_q$ and $\beta_q$ over poles: $( q_1 - q)^2\,=\,0$ and $ (q - q_3)^2 \,=\, 0$. In doing so we obtain
\beq \label{COR2}
\int \, \frac{s\,d \alpha_q \,d \beta_q \,d^2 q_{\perp}}{ 2 \pi i}\,\frac{1}{( q_1 - q)^2}\,\frac{1}{ (q - q_3)^2 }\,\,=\,\,(2 \pi)\,i\,\frac{1}{\alpha_1\,\beta_3\,s}\,\,\,\,\,\mbox{with}\,\,\alpha_q \,=\,-\,\frac{m^2_t}{\beta_3\,s}\,\,\,\mbox{and}\,\,\,\beta_q \,=\,\frac{m^2_t}{\alpha_1\,s}
\eeq
where $m^2_t \,=\,m^2 + (q_{1,\perp} - q_{2,\perp})^2$ or $m^2_t \,=\,m^2 + (q_{3,\perp} - q_{2,\perp})^2$.
>From $\delta\Lb(q_1 - q_2)^2\Rb$ and $ \delta\Lb(q_2 - q_3)^2\Rb$ we know (see section 3.3) that we have
\beq	 \label{COR3}
\int_{\beta_1} \frac{d \beta_2}{\beta_2}\,\int_{\beta_2}\frac{d \beta_3}{\beta_3}
\,\,\,\,\mbox{with}\,\,\,\alpha_1\,=\,\frac{m^2_t}{\beta_2\,s}
\eeq
Multiplying this integral by \eq{COR2} one can see that we have
\beq \label{COR4}
(2 \pi i)\,\int_{\beta_1} \frac{d \beta_2}{\beta_2}\,\int_{\beta_2}\frac{d \beta_3}{\beta_3}\,\,\frac{1}{\alpha_1\,\beta_3\,s}\,\,\propto\,\,\int_{\beta_1} d \beta_2 \,\int_{\beta_2}\frac{d \beta_3}{\beta^2_3}
\eeq
Including the exchange of the extra dilaton one can see that the integral has the following structure
\beq \label{COR5}
(2 \pi )^2\,\,\int_{\beta_1}\beta_2 \,d \beta_2 \,\int_{\beta_2}\frac{d \beta_3}{\beta^3_3}\,\,\,\longrightarrow\,
\,\,\int_{\beta_1}\,\frac{\beta_2}{\beta_2}
\eeq
Therefore, instead of two logarithms from the integration over $\beta_2$ and $\beta_3$ (see \eq{COR3}) we obtain only one logarithm. 

To take these corrections into account we need to change the kernel in \eq{SMP1}, 
\bea \label{COR6}
&&K\Lb z,\bar{z}; z', \bar{z}'\Rb \,\,\longrightarrow\,\,K\Lb z,\bar{z}; z', \bar{z}'\Rb\,\,\,\nonumber\\
&& +\,\,
\Lb\frac{2 \pi}{g_s}\Rb^4\,e^{ - 2 \frac{2 \pi}{g_s}}\,\int\,d q^2 \,\Sigma(q)\,\,\int\,\frac{d z''}{z''^5}\,\frac{d \bar{z}''}{\bar{z}''^5}\,\,
G^2\Lb z, z''; b \Rb\,G^2\Lb \bar{z}, \bar{z}''; b \Rb\,d^2 b\,\,K\Lb z'',\bar{z}''; z', \bar{z}'\Rb
\eea

The numerical estimates show that the second term can reach about $1\,\div 5 \, \%$ of the first one. 
This smallness stems from the ratio $ \int \frac{d x''}{x''}\int G^4 d^2 b/\int G^2\,d^2 b  \approx 3\,10^{-4}$ ($x''=z''/\bar{z}''$). In these estimates we used the simplified version of the kernel given by \eq{SMP7}. 

\section{Conclusions}
In this paper we have demonstrated that the D-instanton induced processes can be a 
dominant source of multiparticle production in the range of the string coupling constant 
$g_s < 1$ but $g_s\,N_c >1$.  The cross section of the instanton-induced multiparticle 
production increases with energy and  the resulting  picture of high energy scattering in 
strongly coupled N=4 SYM looks as follows.  The weakly coupled graviton generates the 
elastic amplitude and the corresponding part of the total cross section, while the 
D-instanton induced interactions of dilatons and axions are reponsible for the 
mutiparticle production processes. This two component picture leads to the structure of 
the high energy interaction which is close to the experimental observations as well as to 
the QCD predictions.

We would like to stress that we are forced to look at the dilaton and axion sector for 
the source of  multiparticle production: as we have discussed above, the AdS/CFT 
correspondence reduces the N=4 SYM in the  strong coupling limit to the action  of 
\eq{SG1} in which the gravitons decouple from dilatons and axions. Since the graviton 
interactions in weakly coupled gravity can only induce elastic interactions (responsible 
for the Newton's law), the only
possibility to describe the multi-particle production is to invoke the dilaton and axion 
sector.
We hope that the proposed here D-instanton--induced mechanism of multiparticle production 
in strongly coupled N=4 SYM provides a resolution of the following dilemma: the 
multiparticle final state (the strongly coupled plasma)
successfully described by this theory seemingly could not be produced in the scattering 
processes dominated by the elastic graviton exchange.

\FIGURE[h]{\begin{minipage}{70mm}
\centerline{\epsfig{file=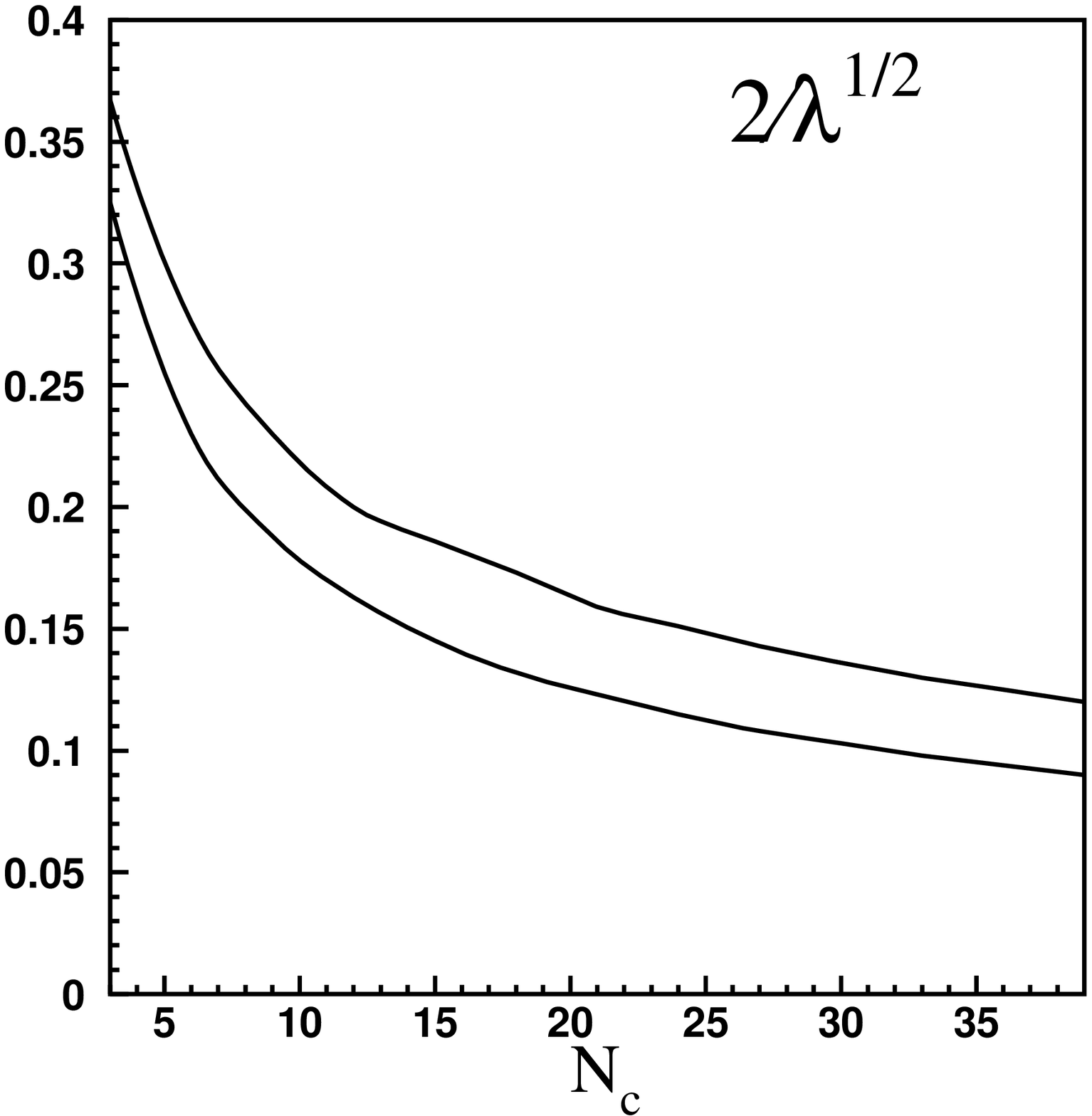,width=60mm}}
\caption{ The maximal and minimum $2/\sqrt{\lambda}$ versus $N_c$ for which $\Delta > 2$ 
and $g_s <1$ while  $g_s \,N_c >1$.}
\end{minipage}
\label{lambda}}

Our estimates of the value of the intercept of the new inelastic Pomeron should be taken 
with caution: the estimate for $z^2_h$ is not reliable and the equation was solved with 
the approximate kernel of \eq{SMP7} instead of one of the full one given by \eq{SMP3}. 
Our numerical calculations show that using the full kernel we obtain $\Delta$ 
approximately two times larger than with the simplified kernel.  In \fig{lambda} we plot 
the dependence the maximal  and minimal values of $2/\sqrt{\lambda}$ versus $N_c$ that 
satisfy the conditions:$g_s\, <\,1, g_s\,N_c>\,1$. For these $2/\sqrt{\lambda}$  the 
value of $\Delta$ is larger than 2
and the instanton induced multiparticle production increases with energy.

It is interesting to compare the pictures of multiparticle production based on QCD 
instanton ladders  at weak coupling \cite{KKL} and on D-instanton ladders in strongly 
coupled N=4 SYM. In the D-instanton picture, we are dealing with the production of
unparticles (bunches of quanta with continuous invariant mass distribution). This is 
analogous
to the production of continuous spectrum of gluons from the decay of the QCD sphaleron.
Another feature of the D-instanton picture is the appearance of a new non-perturbatives 
scale: $z_h$. It plays the role of the typical size of the instanton in QCD vacuum that 
is taken from lattice simulations. It appears that the typical size of QCD instantons and 
the size of D-instantons in our present estimates are quite close to each other: we have 
used $z_h = \sqrt{0.5\ \alpha^\prime} \simeq 0.2$ fm.

D-instanton can be viewed in the string frame \cite{Gibbons:1995vg} as an Einstein-Rosen 
wormhole connecting two asymptotically Euclidean regions of space-time, with the 
Ramond-Ramond (RR) charge flowing down the throat of the wormhole. It describes a 
violation of the conservation of a global charge in physical processes 
\cite{Gibbons:1995vg}. Since in the AdS/CFT dictionary the RR scalar (or the axion, in 
our notations throughout this paper) of supergravity is dual to the $\theta$ angle of N=4 
SYM field theory \cite{Banks:1998nr}, this non-conservation of RR charge describes the 
change of chirality\footnote{In our diagrammatic approach the flow of RR charge is 
signaled by the non-conservation of the number of axions in the D-instanton ladder.}. 
D-instantons thus play a role that is very similar to the role played by the QCD 
instantons that violate chirality conservation.

 It has been predicted \cite{Kharzeev:2004ey} that this violation of chirality
in hot QCD matter would induce the charge asymmetry in high energy heavy ion collisions 
with respect to the reaction plane (i.e. the quark-gluon plasma  would develop an 
electric dipole moment fluctuating on the event-by-event basis). Detailed theory of this 
"chiral magnetic effect" (chiral charge generates an electric current along the direction 
of external magnetic field) has been
developed in Refs. 
\cite{Kharzeev:2007tn,Kharzeev:2007jp,Fukushima:2008xe,Kharzeev:2009pj,Kharzeev:2009ry}. 
The evidence for the chiral magnetic effect from the lattice QCD has been found in 
\cite{Buividovich:2009wi}. The behavior of the effect at strong coupling has been 
explored by holographic methods in \cite{Lifschytz:2009sz,Yee:2009vw,Rebhan:2009vc}. Very 
recently, STAR Collaboration has presented an observation of charge asymmetries in heavy 
ion collisions at RHIC \cite{Abelev:2009uh,Abelev:2009tx}.
Our D-instanton-based approach suggests an important role for topological effects in high 
energy collisions.

~

\section* {Acknowledgements}
The work of  D.K. was supported by the U.S. Department of Energy under Contract No. DE-AC02-98CH10886.
This research of E.L.  was supported in part by BSF grant \# 20004019.

\end{document}